\documentclass[12pt]{article}

\usepackage{authblk}
\usepackage{newtxtext,newtxmath} % keep only once
\usepackage{graphicx}
\usepackage[letterpaper,margin=1in]{geometry}
\usepackage{tabularx}
\usepackage{listings}
\usepackage{setspace}
\usepackage{array,longtable,booktabs}
\usepackage{tikz}
\usetikzlibrary{arrows.meta,calc,patterns,decorations.pathreplacing}
\usepackage{dsfont}
\usepackage{xcolor}                % optional; tcolorbox loads it anyway
\usepackage[most]{tcolorbox}       % <-- required for \newtcolorbox

\usepackage{hyperref}              % usually load last

\usepackage{booktabs, makecell}
\usepackage{threeparttable}
\renewenvironment{abstract}{\quotation}{\endquotation}
\date{}

\makeatletter
\renewcommand{\fnum@figure}{\textbf{Figure \thefigure}}
\renewcommand{\fnum@table}{\textbf{Table \thetable}}
\makeatother
\usepackage{url}
\usepackage{lipsum}
\usepackage{float}

\newtcolorbox{promptbox}{
  enhanced,
  breakable,
  colback=gray!5,
  colframe=black!50,
  boxrule=0.5pt,
  arc=2pt,
  left=1em,
  right=1em,
  top=0.5em,
  bottom=0.5em,
}

\def\scititle{
Who is using AI to code?\\Global diffusion and impact of generative AI}

\title{\bfseries \boldmath \scititle}

\author[1,2]{Simone Daniotti\thanks{Corresponding author: \texttt{daniotti.simone@gmail.com}}}
\author[3,4,2]{Johannes Wachs}
\author[2]{Xiangnan Feng}
\author[2,5]{Frank Neffke}

\affil[1]{\normalsize University of Utrecht}
\affil[2]{\normalsize Complexity Science Hub}
\affil[3]{\normalsize Corvinus University of Budapest}
\affil[4]{\normalsize ELTE Centre for Economic and Regional Studies}
\affil[5]{\normalsize IT:U Interdisciplinary Transformation University}
\begin{document} 
\maketitle
\vspace{-1.2cm}
\begin{abstract} \bfseries \boldmath

Generative coding tools promise big productivity gains, but uneven uptake could widen skill and income gaps. We train a neural classifier to spot AI-generated Python functions in over 30 million GitHub commits by 170,000 developers, tracking how fast ---and where--- these tools take hold. Today, AI writes an estimated 29\% of Python functions in the US, a modest and shrinking lead over other countries. We estimate that quarterly output, measured in online code contributions, has increased by 3.6\% because of this. Our evidence suggests that programmers using AI may also more readily expand into new domains of software development. However, experienced programmers capture nearly all of these productivity and exploration gains, widening rather than closing the skill gap.

\end{abstract}

\clearpage
\noindent

According to proponents, Artificial Intelligence (AI)---in particular generative AI (genAI)---will drastically increase our productivity and revolutionize the way we work. For instance, genAI is expected to complement or substitute humans in an increasing set of tasks \cite{dell2023navigating}. This forces individuals, firms, and policymakers to make important decisions about the use and regulation of genAI under major uncertainty. The stakes are high: genAI has become widely accessible through tools such as ChatGPT or Claude, directly complements human thinking \cite{mollick2024co}, and holds the potential of becoming a general-purpose technology that can solve a wide variety of problems \cite{eloundou2024gpts}. 

Experimental and quasi-experimental evidence so far supports the notion that genAI has transformative potential, showing that genAI leads to increases in productivity and output of individual workers in a variety of jobs \cite{dell2023navigating,brynjolfsson1993productivity,noy2023experimental, cuiEffectsGenerativeAI2024}. Surveys and data reported by Large Language Model (LLM) owners suggest that these technologies are diffusing rapidly \cite{teubner2023welcome,bick2025rapid,humlum2025unequal}. Yet, estimates of the aggregate impact of AI on gross domestic product (GDP) and employment are often modest \cite{acemoglu2025simple,humlum2025large}, suggesting that we are far from having a clear picture of the overall impacts of AI.

We do know that there is significant heterogeneity of adoption which could lead to economic divergence. Although use of genAI is widespread in the working age population, self-reported adoption rates differ markedly across demographics, seniority, work experience, and sectors \cite{bick2025rapid,humlum2025unequal}. Evidence from job ads and firm websites suggests that adoption of genAI varies across geography \cite{andreadis2024local,bearson2025strategic}. If genAI indeed substantially raises productivity, any implied barriers to adoption will have significant consequences for inequality within and across countries \cite{comin2010exploration}. Historically, macro-level productivity effects of general-purpose technologies, such as steam engines, electricity, and computers, have taken long to materialize \cite{david1990dynamo,brynjolfsson1993productivity, brynjolfsson2021productivity, juhasz2024technology}. Together, this leads to substantial uncertainty about the impact of genAI today. 

Resolving this uncertainty requires accurately determining adoption rates, intensity of use, and productivity effects at a global level. Surveys demonstrating demographic and sectoral heterogeneities in genAI adoption focus on single countries \cite{bick2025rapid,humlum2025unequal}. Previous work comparing AI adoption in different countries using survey data finds evidence of differences within and between countries \cite{calvino2023portrait}, but differences in sample weighting and analysis periods of the surveys limit our ability to directly compare observed rates. In the context of genAI, respondents may under-report usage, especially at work, to avoid judgment \cite{ling2025underreporting, Almog2025JMP}. Nevertheless, surveys provide a valuable resource for understanding adoption patterns. Similarly, randomized controlled trials (RCTs) \cite{peng2023impact,cuiEffectsGenerativeAI2024,dell2023navigating,brynjolfsson1993productivity,noy2023experimental} and natural experiments \cite{hoffmann2024generative,yeverechyahu2024impact,song2024impact} are indispensable because they measure causal effects of genAI adoption by design. However, they typically consider individuals as ``treated'' whenever they have access to genAI tools, leaving the extent to which treated individuals used genAI during the experiments unknown. Moreover, surveys and experiments tend to observe individuals over short time periods, which limits our ability to know the dynamics of adoption and to observe effects of adoption that materialize more slowly. 

To begin to address these gaps, we ask if we can measure the adoption and use of genAI in another way at the level of individuals whom we can track over longer periods of time. If so, what do such measures tell us about the rate of adoption of genAI? Does this differ across countries and demographics? How does genAI impact the outputs individuals produce, and how do individual characteristics such as experience moderate these effects?

To answer these questions, we study genAI use at a fine-grained level in one of its main domains of application: software development, an important and high-value sector \cite{aum2024software,juhasz2024software} that is uniquely exposed to genAI \cite{handa2025economic, hoffmann2024generative,peng2023impact,del2024large}. To do so, we design and implement a machine learning classifier to identify code written with substantial AI assistance in over 30 million software developer contributions, also known as commits, to open-source Python projects on GitHub. To train this classifier, we assemble a custom training set, combining existing sources with a procedure that generates synthetic training data. This allows us to analyze shifting patterns of AI-generated code at a granular level. We leverage this novel source of micro-data to study how quickly the use of genAI in coding diffuses in six major countries, how this diffusion relates to demographic characteristics, and how it affects programming activity in a sample of over 100,000 US software developers.

\section*{Detecting AI generated code}

To collect a large dataset of coding activity, we gather all commits by about 100k US GitHub users to Python-based open-source projects, recursively cloning all GitHub directories related to each project. Next, we add commits from a random sample of 2,000 programmers per year for each of five other major countries in software development: China, France, Germany, India and Russia.  We then analyze these commits to assess the prevalence of AI-generated code (see Materials and Methods).

\begin{figure}[H]
    \centering
    \includegraphics[width=\textwidth]{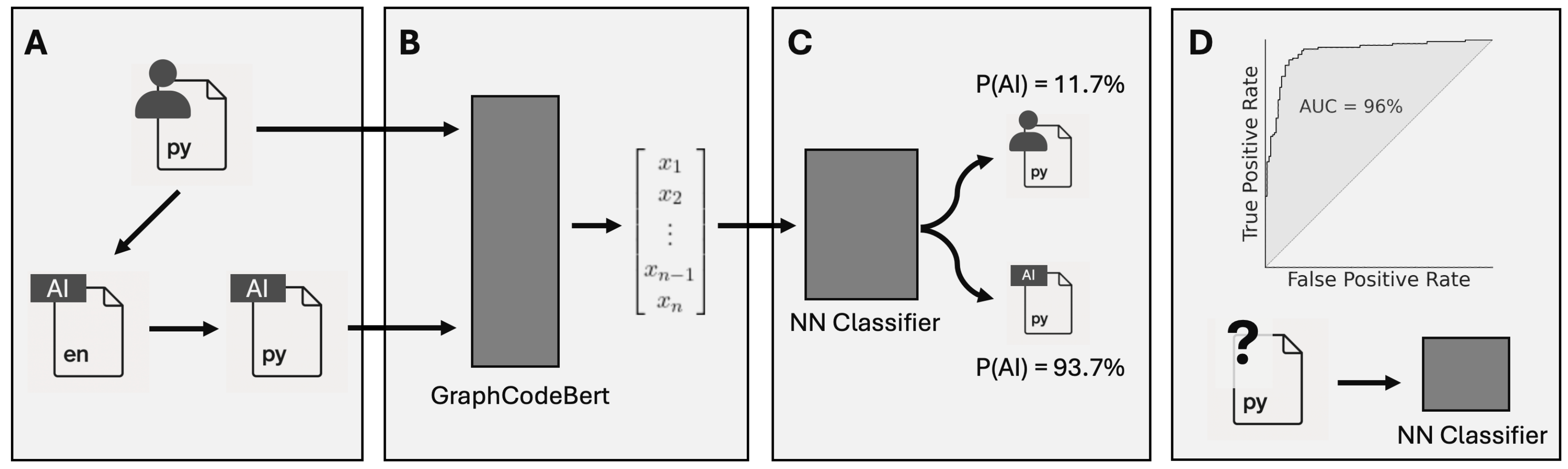}
    \caption{Classifying code from functions written in the Python programming language as human or AI generated. A) Using a collection of human generated code, we ask one LLM to describe the code in English, then another to implement that description as a Python function. B) We vectorize the resulting code using GraphCodeBert, an embedding method that uses a code’s tokens, comments, and variable graph flow. C) We train a neural network classifier combining GraphCodeBert with a classification head to predict the human/AI labels. D) We evaluate the classifier on out-of-sample data and apply it to a large database of unlabeled Python functions.}
    \label{fig:1}
\end{figure}

Figure~\ref{fig:1} describes how we classify these code contributions as either human or AI generated.
We limit this analysis to blocks of code that represent functions to focus on a fine-grained, self-contained, yet substantial unit of code. We first construct a ground truth dataset (Fig.~\ref{fig:1}A),  collecting Python functions of which we are certain they were written by a human programmer. To do so, we take functions written in 2018, as they predate the release of capable genAI models.
Because programming styles evolve over time, we add functions created in later years from the HumanEval datasets for the years 2022 and 2024. To add a dataset of similar functions but written by genAI we apply a two-step procedure. 
First, for each human-written function we ask one LLM to describe the function in English, specifying the type of input and output of the function.
Second, we feed this text to a second LLM and request the model to generate a function based on this description. Our use of two different LLMs ---unlike previous approaches \cite{yeUncoveringLLMGeneratedCode2024}---   avoids creating unnecessarily strong correlations between human code and its transcription, while ensuring that the (synthetic) AI-generated functions in our training data are close in functionality to the original human-written functions.

We then train a machine learning classifier on this dataset. Following \cite{nguyenGPTSnifferCodeBERTbasedClassifier2024}, we transform each  function using \emph{GraphCodeBert}, a pre-trained language model for code that embeds a function into a high-dimensional vector space using its tokens, comments, and the dataflow graph of its variables \cite{guoGraphCodeBERTPretrainingCode2021}. The resulting vectors are fed into a classifier to determine whether a given function was written by a human or by genAI (Fig.~\ref{fig:1}B-C). 

\section*{Results}

The classifier performs remarkably well, achieving an out-of-sample ROC AUC Score of 0.96 (Fig.~\ref{fig:probabilities}D) and Average Rate of True positives of 0.95. We apply this classifier to 5 million functions extracted from 31 million contributions to Python projects from the beginning of 2019 to the end of 2024 for the full population of US-based users and the sampled  users in the five other countries (Fig.~\ref{fig:1}D). In the Materials and Methods, we show that the classifier also correctly identifies code generated by more recent LLMs introduced after our data collection ended, as well as code produced in real-world interactions with LLMs, albeit with somewhat lower accuracy. Retraining the classifier on code produced by these newer LLMs further improves performance. 

\begin{figure}[H]
    \centering
    \includegraphics[width=0.7\textwidth]{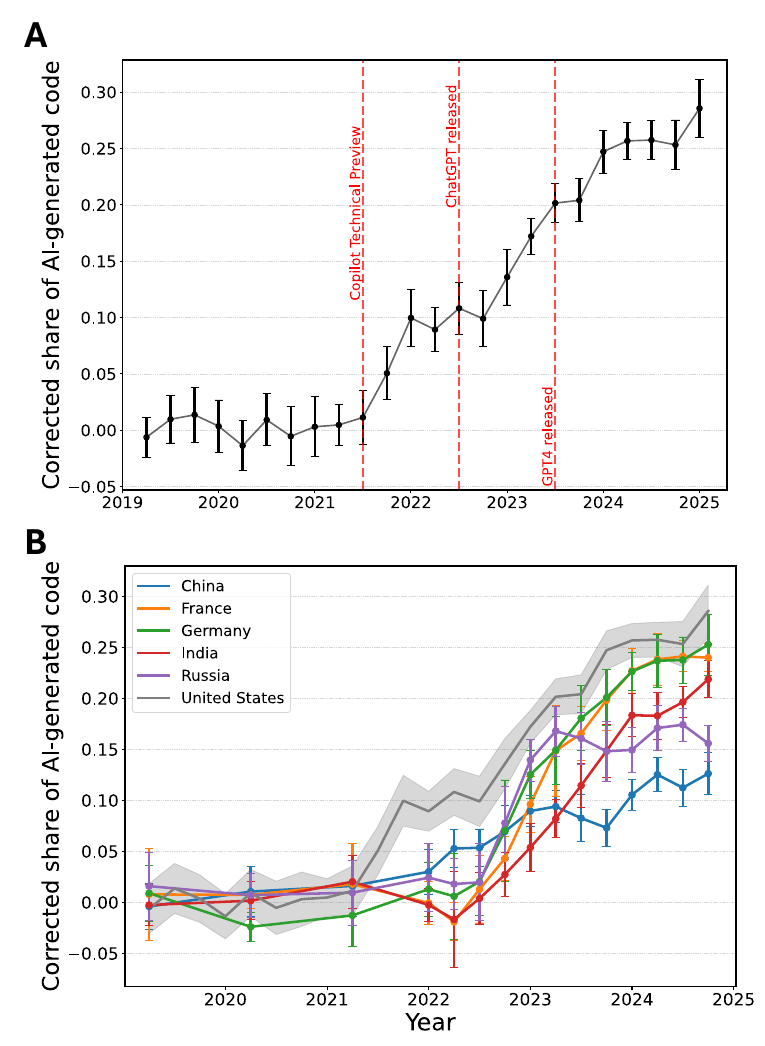}
    \caption{\textbf{Share of AI-generated Python functions over time.} \textbf{A}: share of Python functions that were created or substantially changed by GitHub users in the United States. Vertical lines depict 95\% confidence intervals. The plot reveals abrupt shifts in adoption coinciding with key AI-related events: the release of GitHub Copilot Preview, the public launch of ChatGPT, and the second wave of LLM releases (GPT4 and related models).  \textbf{B}: adoption in China, France, Germany, India and Russia (note that in China, GitHub competes with the alternative collaboration platform, Gitee\cite{gortmaker2024open}). We sampled 2,000 random programmers per country-year. The US curve is replicated from panel A as point of reference. The US lead the early adoption of genAI, followed by European nations such as France and Germany. From 2023 on, India rapidly catches up, whereas adoption in China and Russia progresses more slowly.}
    \label{fig:adoption}
\end{figure}

Figure~\ref{fig:adoption}, panel A plots the AI adoption trajectory for US developers. Adoption rates sharply increase following major AI advancements, including the launches of Copilot, ChatGPT, and second generation LLMs. Panel~B compares the US to the five other major countries we cover in the global race toward AI adoption. This shows that the US took an early lead, which it has managed to maintain ever since. About 29\% of Python functions in the US were generated by AI by the end of 2024, closely followed by 23/24\% for Germany and France. India draws close at 20\%, after having initially lagged in adoption. In contrast, Russia and China have so far remained late adopters. 

\begin{figure}[H]
    \centering
    \includegraphics[width=0.7\linewidth]{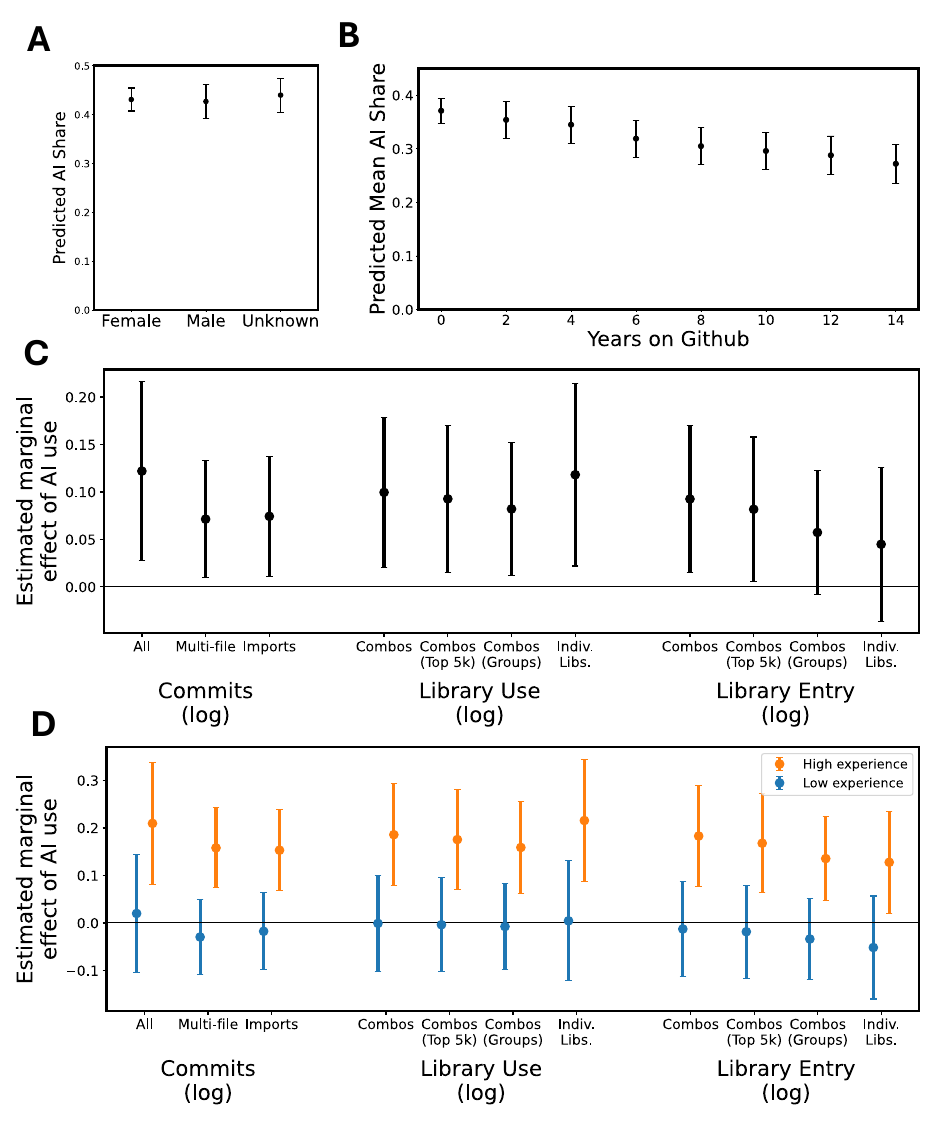}
    \caption{A) Intensity of genAI use by gender inferred from GitHub display names (US, 2024). B) Intensity of genAI use by user tenure (US, 2024). C) Estimated effect of genAI use on user activity from a user-quarter panel regression with user and quarter fixed-effects. GenAI use is associated with increased commit activity across all commits, multi-file commits (``Multi-File'') that navigate project interdependencies, and commits adding library imports (``Imports''), which we interpret as adding new features. GenAI is also associated with wider ranges of individual libraries (``Indiv. Libs'') and library combinations (``Combos''), and increased experimentation with new libraries or combinations. Results hold subsetting on the 5,000 most common library combinations (``Combos (Top 5k)'') and coarsened library categories (``Combos (Groups)''). Error bars: 95\% confidence intervals (standard errors clustered by user).}
    \label{fig:3}
\end{figure}

Focusing on the full population of US developers, we find that AI adoption rates drop with the number of years that developers have been active on GitHub. Fig.~\ref{fig:3}B shows that whereas the most experienced developers use genAI in 27\% of their code, programmers who have just joined the GitHub platform use these tools for 37\% of code. In contrast, using (self-reported) first-name-based gender inference algorithms, we find no difference between men and women (Fig.~\ref{fig:3}A).

To assess how genAI impacts the quantity and nature of code that programmers produce, we rely on regression models with user and quarter-of-year fixed effects. This compares the output ---  in terms of quarterly number of commits ---  of the same programmer at different points of AI adoption, controlling for economy-wide trends. These models, summarized in Fig.~\ref{fig:3}C, suggest a substantial impact of genAI on developer productivity. We find consistent effects across different sets of commits: all commits, commits that modify multiple files (which typically require navigating complex dependencies across scripts), and commits that add new libraries or library combinations (which typically introduce new functionality to scripts). Moving from 0 to 29\% genAI usage—the estimated US adoption rate by the end of 2024—is associated with a 3.6\% increase in commit rates across all these commit types. However, these associations with user productivity are fully driven by experienced users, for whom a 29\% adoption rate would imply a 6.2\% increase in commit rates (Fig.~\ref{fig:3}D). In contrast, we observe no  statistically significant effects among inexperienced users.

Apart from increasing activity rates, AI adoption is also associated with increased experimentation with new libraries and combinations of libraries, which \cite{fang2024novelty} interpret as signs of innovation. Because libraries often focus on specific types of functionality --- such as visualization, natural language processing, web interactions, or database operations --- this suggests that genAI helps programmers expand their capabilities to new domains of software development. 
At average end-of-2024 AI use rates for US developers, our models predict that developers will implement 2.7\% more new combinations of libraries. Results are robust to variations in how we determine new library introductions. In particular, effects are unlikely due to esoteric libraries (``AI slop''): findings do not change much if we only use the 5,000 most common libraries or if we group libraries into 124 coarse categories first. Moreover, Fig.~\ref{fig:SI_measurementerror} of the SI shows that these effects, as well as the earlier productivity effects, are likely lower bounds because errors in the measurement of users' AI adoption rates bias each of these estimates downwards.

\section*{Discussion}

We set out to measure the use of genAI at the micro-level, in order to study its diffusion and its consequences at a global scale. Focusing on the software development labor force, we demonstrated how genAI has diffused and how this has affected the quantity and nature of code that programmers produce. To do so, we developed a new genAI classifier to identify AI-generated functions in GitHub commits. Applied to  a large dataset covering software development activity across major countries, we document noticeable growth spikes in genAI-generated code soon after key genAI releases. Yet, we also observe significant differences between countries. Corroborating existing studies \cite{bick2025rapid,humlum2025unequal}, our estimated adoption rates are higher among less experienced programmers but unlike most previous work, we find no significant differences between men and women. 

We also find that genAI reshapes both the volume and nature of programming work. Using within-developer variation—comparing the same programmer before and after adopting genAI—we show that AI adoption substantially increases output. Developers using genAI are also more likely to incorporate novel combinations of software libraries into their code, suggesting they venture into new technical domains \cite{fang2024novelty} using unfamiliar building blocks \cite{eghbal2016roads}. However, both productivity and exploration gains concentrate almost exclusively among more experienced developers.

How do our findings compare to prior findings? Our estimates of the most recent adoption rates in the US of around 29\% are remarkably similar to adoption rates claimed for coding work at Microsoft \footnote{\href{https://cnb.cx/3YorqDQ}{https://cnb.cx/3YorqDQ}} and Amazon \footnote{ \href{https://www.nytimes.com/2025/05/25/business/amazon-ai-coders.html}{https://www.nytimes.com/2025/05/25/business/amazon-ai-coders.html}}. This shows that, despite our focus on code from open source Python libraries, our results closely align with estimates of adoption rates from other contexts and may generalize beyond the specific setting of this study. 

Unlike most other studies, our design allows us to compare adoption rates across countries. Here, we find a clear and sustained lead by US programmers. Use of LLMs may be lower in countries like China and Russia because providers such as OpenAI and Anthropic block access (supply side) and censorship limits local use (demand side), even though many users connect using VPNs~\cite{bao2025there}. However, other major countries are quickly catching up, eroding the US' first-mover advantage. Another aspect that sets our study apart is that existing literature typically focuses on \emph{access} to genAI --- yielding reduced-form estimates of the causal effect of the so-called intention-to-treat, not of genAI itself --- or usage in controlled experimental settings. In contrast, our approach allows us to quantify the intensity with which the technology is used in real-world work activities. Finally, we note that our cross-country evidence on genAI use complements firm-level survey work on broader AI adoption which extends back to before the genAI era \cite{calvino2023portrait}; while levels are not directly comparable, both perspectives document persistent cross-country gaps in AI use.

When it comes to the productivity effects of genAI, our estimates are generally smaller than those found in RCTs \cite{paradis2025aiimpact,cuiEffectsGenerativeAI2024} and studies exploiting natural experiments \cite{hoffmann2024generative,song2024impact}. In robustness checks (SI section~\ref{sec:SI_regression}) we study whether nonlinearities or threshold effects in the benefits of genAI adoption can explain such discrepancies, but find little significant evidence for this hypothesis. A more promising explanation is measurement error, which is likely to bias effect estimates downwards. In line with this, Fig.~\ref{fig:SI_measurementerror} of the SI shows that our effect estimates increase substantially if we correct them for measurement error. Moreover, we show that effects concentrate in experienced programmers, while junior developers seem not to benefit much from genAI. The higher effect estimates in prior literature may therefore also reflect differences in the populations and complier samples they studied.  

There are several limitations to our study. First, our analysis focuses on software development. Although this limits its scope, work in this sector is uniquely amenable to quantitative analysis at a level of granularity that is required to study how AI affects workers and their tasks. Within software, we focus only on Python-based open-source contributions. While Python is a widely used language, adoption patterns may differ in other programming ecosystems. We argue that estimates derived from open‑source Python code on GitHub are economically meaningful, because open source software (OSS) underpins most commercial stacks and carries large measured value \cite{eghbal2016roads,hoffmann2024value}. GitHub’s central role in collaboration, networking, and signaling further ties our evidence to professional activity \cite{dabbish2012social,abou2025career}. Finally, that our estimates of AI use in the US line up closely with reported AI use at leading US firms mentioned above increases our confidence in the external validity of our findings.

More generally, we also do not account for potential externalities between coworkers or heterogeneity in productivity across firms, all of which may be relevant factors in how genAI affects programming activity. Beyond firms, our geographic analysis is limited to a subset of countries and it would be important to widen the analysis to include countries at different income levels. In the specific case of China --- where the programming community also relies on an alternative collaboration platform, Gitee \cite{gortmaker2024open} --- there is some additional risk that our focus on GitHub projects distorts estimates. Finally, when it comes to the effects of genAI, there are many other ways to evaluate the productivity of programmers that heed more attention to code quality, from tracking how issues get resolved and code merges to the  implemented test coverage. While feasible in principle, such analysis requires new data collection and careful design of metrics. We therefore leave questions of the effect of genAI on code quality for future research.

How much value has genAI created in coding? While hard to answer definitively, our study offers some important pieces of this puzzle. Based on an analysis of detailed task surveys and wage statistics for about 900 different occupations, we estimate that the US spends between 637 and 1,063B USD on labor costs related to coding activities per year (SI, section \ref{sec:SI_wagesum}). Assuming our estimated diffusion rates of 29\% by the end of 2024 (based on open-source Python contributions) are representative of code in general, the annual value generated by genAI coding assistants in the US would depend on how much they increase productivity. Using our own, conservative, baseline estimates, genAI would have increased the volume of commits by 3.6\%. Assuming these commits reflect valuable code contributions, our calculation implies that genAI generates $23–$38B USD of additional code per year. This estimate treats productivity gains as similar across programming languages. In a more conservative scenario, where productivity effects outside Python are negligible, the value of genAI would drop to about 17\% of this figure (approximately $4–$6B), using estimates of Python’s share of GitHub code \cite{github_language_stats}.

By contrast, various lab experiments \cite{ peng2023impact,paradis2025aiimpact} and field experiments \cite{cuiEffectsGenerativeAI2024}  in software development all yield substantially larger causal effects of genAI on task completion times --- arguably a more relevant quantity to track than commit volumes. Averaging across such studies (see Materials and Methods for details) yields an estimated 6.0\%-15.7\% increase in productivity at a 29\% adoption rate, translating into a range of 38-167B USD for the direct impact of genAI on US coding activities. However, these estimates ignore that genAI may also lead to a reduction in the market price of code. Because this yields cost savings for consumers of code, while reducing profits for suppliers (i.e., programmers),  factoring in such general equilibrium effects further widens the range of possible outcomes  (SI section~\ref{sec:SI_geneq}). In the Materials and Methods section we show that this would mostly affect the upper bound of our estimates, with lower bounds all but unaffected.  
The upshot of these back-of-the-envelope calculations is that, although the total value of genAI to the US economy is uncertain, it is most likely substantial, on the order of at least tens of billions of USD.  

Given that genAI has diffused quickly beyond the US, global cost savings would be larger still, even if we confine ourselves to the software sector. Moreover, we are currently still in the early phases of the diffusion curve of what looks to be a new general purpose technology~\cite{eloundou2024gpts}. Historically, early-stage productivity effects of general purpose technologies have been hard to identify because it takes time to integrate them into firm level workflows and procedures, train individuals and amass the complementary assets needed to fully exploit their potential. Based on this, we find ourselves on the bullish side of the debate when it comes to the productivity effects of genAI.

Our results on such effects and the heterogeneous diffusion of genAI raise important questions for policymakers and researchers. We need to understand the barriers of adoption to AI: are these similar to prior radical innovations~\cite{frenken2023new} or is this time different? Additionally, these barriers need to be understood not only at the individual level, but also at the firm, regional, and national levels. Our study takes a first step toward answering such questions.

Moreover, given the wide dispersion in productivity across programmers \cite{sackman1968exploratory,bryan1994not,betti2025dynamics} and our finding that benefits accrue to more experienced coders only, future research should explore how AI adoption affects developer activity at the upper tail of elite programmers, where the most significant breakthroughs and innovations are likely to occur. Finally, our study exclusively focused on programming tasks. Yet, one study of elite software developers suggests that access to genAI leads to a shift from managerial tasks to coding \cite{hoffmann2024generative}, suggesting that an important margin along which productivity effects materialize is shifts in the task composition of software developer jobs. 

The nature of work often changes with the introduction of new technologies. Understanding these changes is especially difficult when the innovation in question is radical \cite{frenken2023new}, such as the spinning jenny,  transistors, or robots, and at the same time pervasive \cite{freeman1988structural}. The uncertainty of the effects of genAI on work and labor markets is reflected in the wide range of attitudes researchers and policymakers take towards it, ranging from utopian to skeptical and outright apocalyptic. These attitudes are formed in a fast-moving context, and are based on incomplete evidence on the adoption and effects of AI. The findings in this study provide better evidence of how genAI is used in a large, important, and highly exposed sector of the economy, as well as a way to monitor this in real-time going forward. Applying our AI detection classifier to millions of code contributions made over a six-year period, we can confirm that AI adoption is fast, but heterogeneous across countries and individuals. Moreover, AI adoption is associated with increased activity rates in online software development collaborations.

However, one of the most surprising findings is the fact that genAI increases experimentation with new libraries, suggesting that genAI allows users to advance faster to new areas of programming, embedding new types of functionality in their code. This corroborates prior findings \cite{doshi2024generative} that genAI increases individual innovation, pushing individuals' capabilities in terms of the use of new combinations of libraries. However, again only experienced users seem able to leverage genAI in this way, with important consequences for programmers' ability to master new coding skills in the presence of genAI.

%%%%%%%%%%%%%%%% ACKNOWLEDGEMENTS %%%%%%%%%%%%%%%

\section*{Acknowledgments}
We thank Ulrich Schetter, Hillary Vipond, Andrea Musso, and participants of the ANETI Brownbag seminar for helpful comments. We thank M\'arton Salamon for valuable research assistance.
\paragraph*{Funding:}
F.~N., S.D., J.W. and X.F. received financial support from the Austrian Research Promotion Agency (FFG) in the framework of the project ESSENCSE (873927), within the funding program Complexity Science. JW also acknowledges financial support from the Hungarian National Scientific Fund (OTKA FK 145960).
\paragraph*{Author contributions:}
S.D. and F.N. conceptualized the research. S.D. implemented the primary method. S.D. and J.W. collected the data. X.F. collected data and estimated the volume of the U.S. programming-related wage sum. All authors analyzed the data. F.N. and J.W. led in drafting the manuscript. All authors contributed to the writing of the manuscript.

\paragraph*{Competing interests:}
There are no competing interests to declare.
\paragraph*{Data and materials availability:}
Code and data to replicate our analyses are available here: \href{https://drive.google.com/drive/folders/1h-KTmHNRzFD_vTFeHzgUUALhHPg2y9cW?usp=sharing}{this link}

%%%%%%%%%%%%%%%% SUPPLEMENT TITLE PAGE %%%%%%%%%%%%%%%
\clearpage
\setcounter{section}{0}
\renewcommand{\thesection}{S\arabic{section}}

\setcounter{figure}{0}
\renewcommand{\thefigure}{S\arabic{figure}}

\setcounter{table}{0}
\renewcommand{\thetable}{S\arabic{table}}

\setcounter{equation}{0}
\renewcommand{\theequation}{S\arabic{equation}}

\setcounter{page}{1}
\renewcommand{\thepage}{S\arabic{page}}
\begin{center}
\section*{Supplementary Materials for\\ \scititle}

Simone~Daniotti$^{\ast}$,
Johannes Wachs,
Xiangnan Feng,
Frank Neffke\\
\end{center}

\subsubsection*{This PDF file includes:}
Materials and Methods\\
Supplementary Text\\
Figures S1 to S9\\
Tables S1 to S28\\

%%%%%%%%%%%%%%%% MATERIALS AND METHODS %%%%%%%%%%%%%%%

\section{Materials and Methods}

\section*{Data}

Data and code to reproduce the results in this paper are available at
\href{https://github.com/SimoneDaniotti/ai_in_code_daniotti}{this link}. The datasets contain the following columns: anonymized user ID, anonymized ID of the modified function, anonymized project name, anonymized commit ID, estimated (corrected, see section~\ref{sec:SI_bayesiancorrection}) AI probability, and the user's self-reported country.

\begin{figure}[H]
    \centering
    \includegraphics[width=\textwidth]{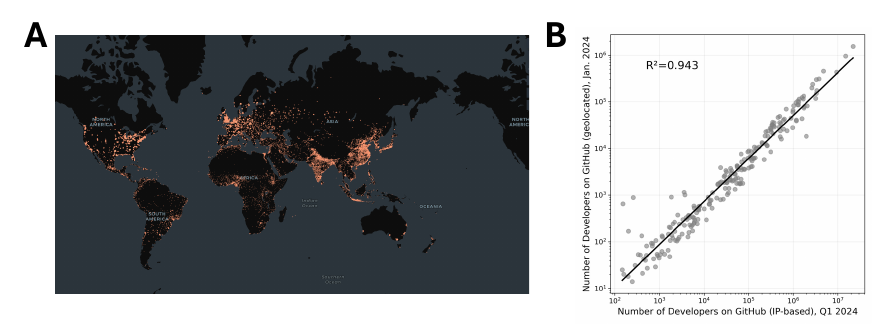}
    \caption{\textbf{Location of GitHub users.} \textbf{A}: Self-reported locations of GitHub users. \textbf{B}: Number of users in each country based on self-reported locations against IP addresses.}
    \label{fig:positions}
\end{figure}

Our data collection proceeds as follows. First, we collect all users who have made a commit to public repositories on GitHub using the GHArchive dataset hosted on Google BigQuery, focusing on the period 2019-2024. Then, we geolocate GitHub users using self-reported locations in GitHub profiles obtained through the \emph{GraphQL} API~\cite{GitHubGraphQL}. Fig.~\ref{fig:positions} of the SI shows that the distribution of users across countries derived from self-reported data closely matches the same distribution derived from users' IP addresses as provided in the GitHub Innovation Graph dataset (\url{https://innovationgraph.github.com/}). 

Next, we collect commits by users that report US locations to projects that use Python as their programming language. To do so, we select all users that are active in such projects in a given year. Next, in each year, we  recursively clone all GitHub directories related to these users and obtain all commits to projects where the user makes over 3 commits in the year. 
Finally, we analyze these commits using the \emph{PyDriller} tool~\cite{spadiniPyDrillerPythonFramework2018}. 
To manage computational expenses, AI usage in the other five countries of  Fig.~\ref{fig:adoption}\textbf{B} is based on random samples. We draw these samples from populations that are created analogously to the one for the US. From each population, we sample 2,000 programmers per country per year. This yields  a total of 70,000 user-year observations. 

\section*{Detecting Generative AI in code}

To detect the use of genAI, we focus on self-contained code blocks that  perform well-defined tasks: functions. For each function, we determine how many lines were modified in a commit. We then keep only those functions in which modifications occurred in over $80\%$ of lines of code.

\subsection{Training data}
The ground truth data on which we train our supervised model for detecting AI-generated Python functions combines multiple sources. We start by collecting functions created by human coders from three different datasets. The first dataset randomly samples Python functions from GitHub that were created before the introduction of genAI coding tools. In particular, we sample python functions created in 2018. The second and third datasets are HumanEval~\cite{chenEvaluatingLargeLanguage2021a} and  HumanEval-X~\cite{zhengCodeGeeXPreTrainedModel2023a}, both of which were originally created to evaluate and measure functional correctness of code. All the three dataset combined contain almost 4,000 human-written Python functions. Adding the latter two datasets ensures that the human-made Python functions we use to train our models include examples created in different years, including years in which genAI tools had become more widely available. 
Next, inspired by~\cite{yeUncoveringLLMGeneratedCode2024}, we generate a synthetic dataset of Python functions written by different LLMs, using GPT3.5-turbo (50\%), GPT-4o-mini (30\%) and GPT-4 (20\%). To do so, we create synthetic clones of the human-written functions described above, using an LLM chain that combines two LLMs, mimicking recent LLM Agent \href{https://langchain-ai.github.io/langgraph/agents/tools/}{tools}.  Each LLM performs a different task. The first LLM is prompted to describe a given human-made function in terms of its functionality and the structure of the required input and generated output. The second LLM is asked to read this description, and to then generate a function that accomplishes the same task. The exact prompts and an example of the output are listed in Table~\ref{tab:llm-prompts} of the SI.
The total size of the training set amounts to around 8,000 functions.

\subsection{Detection model}
Our genAI detection model relies on open source components and is set up to efficiently scale to analyze millions of Python functions.
We chose a state-of-the-art technique based on CodeBERT~\cite{fengCodeBERTPreTrainedModel2020}.
Our approach resembles that of GPTSniffer~\cite {nguyenGPTSnifferCodeBERTbasedClassifier2024}. 
In this paper, the authors train a CodeBERT model to detect AI-written Javascript programs. 
We instead aim to detect AI-written Python functions. 
To do so, we build a classifier on top of a newer and more advanced version, GraphCodeBERT~\cite{guoGraphCodeBERTPretrainingCode2021}, which  is better able to capture and understand patterns in code. We add a linear layer to the GraphCodeBERT model to perform a classification task. 
Finally, we use the training data  to finetune all parameters for optimal model prediction, including to retrain the embedding weights to optimize results.

\begin{figure}[h!]
    \centering
    \includegraphics[width=0.95\textwidth]{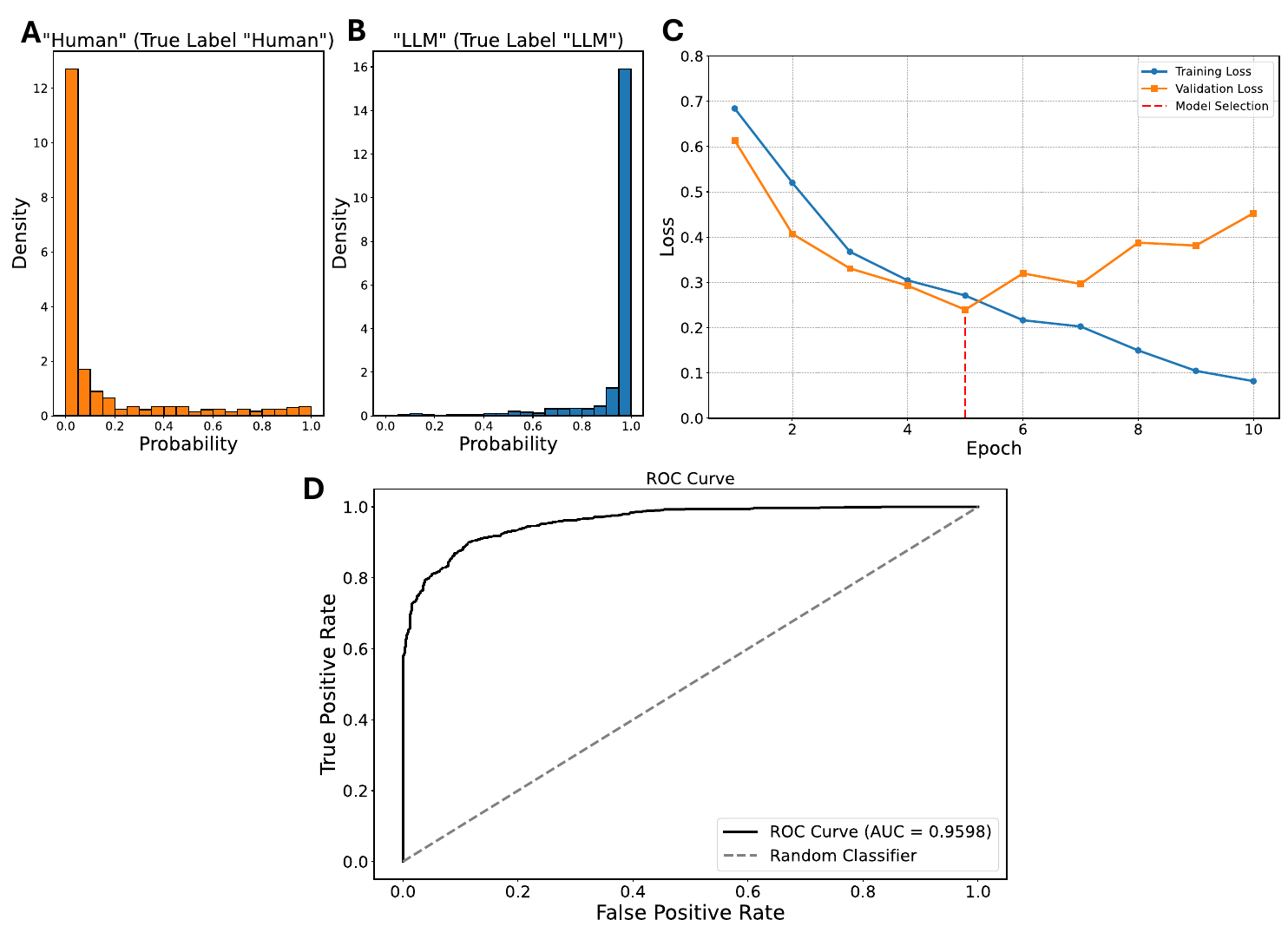}
    \caption{\textbf{Detector Prediction Test.} Evaluation of the trained detector on a test set. \textbf{A}: predicted probability that code was AI-generated for human-generated functions. \textbf{B}: predicted probability that code was AI-generated for AI-generated functions. \textbf{C}: Loss curve during the training of the detection model.
    \textbf{D}: ROC Curve of the  classifier.}
    \label{fig:probabilities}
\end{figure}

Tokenization was performed using GraphCodeBERT’s tokenizer, setting a maximum sequence length of 512 and applying padding and truncation to maintain consistency across inputs. As a loss function, we use cross-entropy loss (\texttt{torch.nn.CrossEntropyLoss}), which is well-suited for classification problems and commonly used in RoBERTa-based classification models. As shown in Fig.~\ref{fig:probabilities}, in our case it learns to effectively  differentiate between AI-generated and human-written code. Since the model returns raw logit values, the model internally applies a softmax operation, comparing the predicted probability distribution with the ground truth labels.

To train the model, we split our ground truth data into training and evaluation sets using an 80/20 ratio with a fixed random seed. We then use the Hugging Face Trainer API to handle training, evaluation, and optimization. Our training configuration includes 10 epochs, a batch size of 32 per device for both training and evaluation and the AdamW optimizer (\texttt{adamw\_hf}) with a learning rate of \( 1\text{e-}5 \) and weight decay of \( 0.005 \). We set \texttt{warmup\_steps=1000} to help stabilize learning in the early phases. Logging occurs once every 100 steps, with evaluation and model check-pointing at each epoch. In the end, only the best-performing model on the evaluation set is retained. For reproducibility, we use a fixed data seed (\texttt{seed 365}). As shown in Fig.~\ref{fig:probabilities}, the model manages to effectively identify AI generated code in our ground truth data, reaching an out-of-sample ROC AUC Score of $0.96$, Average Precision Score~\cite{scikit-learn-metrics} of $0.9685$ and Average true positive rate of 0.95 (F1 score $0.8911$ with a 0.5 threshold).

\begin{table}[ht]
\centering
\begin{tabularx}{\textwidth}{|l|X|}
\hline
\textbf{Role} & \textbf{Prompt} \\
\hline
System & 
You are a python expert programmer. \\
\hline
User (Generate description) &
This is a python script in markdown. 
Describe the task or tasks this script is solving,
explaining the input and the output specifications for each function.\\
\hline
User (Code from description) &
This is the description of a python script.
Based on the description, write a full code that fulfills that task/tasks. 
The python script should be organized in a single markdown block. 
Please return only the code, do not return any clarifications before or after the code.\\
\hline
\end{tabularx}
\caption{System and User Prompts for Synthetic Dataset generation.}
\label{tab:llm-prompts}
\end{table}

\subsection{Estimating AI usage rates}\label{sec:SI_bayesiancorrection}

In Fig.~\ref{fig:adoption} and in the regression analyses, we study the diffusion and effects of genAI by quantifying the probability that a piece of code was written by AI: $P(A=1)$. Our data allow us to estimate a different probability: the probability that our model detects the use of AI in a function: $P(D=1)$. Using the law of total probability, we can write the latter probability as:
\begin{align}\label{eq:law_of_tot_prob}
    P(D=1) &= P(D=1|A=1)P(A=1) + P(D=1|A=0)P(A=0) 
\end{align}
or
\begin{align*}
    P(D=1) &= P(D=1|A=1)P(A=1) + P(D=1|A=0)\left(1-P(A=1)\right) 
\end{align*}

We can estimate some of these terms using our ground truth data set:
\begin{itemize}
    \item  $P(D=1|{A}=1)$: $\hat{d}^{AI}_{GT}$, estimated probability that AI-generated code is detected to be AI  in the ground truth data (true positive rate)
    \item $P(D=1|{A}=0)$: $\hat{d}^{hum}_{GT},$ estimated probability that human-written code is  detected to be AI in the ground truth data (false positive rate)
    \item $P(D=1)$: $\hat{d},$ estimated probability of AI detection (observed quantity) 
    \item $P({A}=1)$: $\hat{y},$ estimated AI usage rate (quantity of interest)
\end{itemize}

Using this notation, we can write eq.~(\ref{eq:law_of_tot_prob}):
\begin{align*}
     \hat{d} &= \hat{d}^{AI}_{GT}\hat{y} + \hat{d}^{hum}_{GT}\left(1-\hat{y}\right) \\
     &= \hat{y}\left(\hat{d}^{AI}_{GT} - \hat{d}^{hum}_{GT} \right) + \hat{d}^{hum}_{GT} 
\end{align*}
and rearrange terms to arrive at:
\begin{align}\label{eq:rearrange}
     \hat{y} &=  \frac{\hat{d} - \hat{d}^{hum}_{GT}}{\hat{d}^{AI}_{GT} - \hat{d}^{hum}_{GT}},  
\end{align}

We can use this equation to correct the estimated AI adoption probabilities from our AI detector for miss-classification errors. We use this quantity throughout the paper as the estimated AI adoption rate in a given sample of functions.
In doing so, we allow true and false positive rates to differ across countries, using country-specific correction parameters  $\hat{d}^{AI}_{GT}$ and $ \hat{d}^{hum}_{GT}$.
As shown in  Table \ref{tab:false_positive_rates}, such differences turn out to be modest and most pronounced in false positive rates. 

\begin{table}[ht]
\centering
\begin{tabular}{l c c}
\toprule
\textbf{Country} & False Positive Rate & True Positive Rate \\
\midrule
United States & 0.2321 & 0.9550 \\
China         & 0.2405 & 0.9521 \\
Germany       & 0.2397 & 0.9516 \\
India         & 0.2296 & 0.9520 \\
Russia        & 0.2572 & 0.9617 \\
France        & 0.1989 & 0.9519 \\
\bottomrule
\end{tabular}
\caption{Country-specific estimates of the false positive rate, \(\hat{d}^{\text{hum}}_{\text{GT}}\), and true positive rate, $\hat{d}^{AI}_{GT}$.}
\label{tab:false_positive_rates}
\end{table}

A consequence of the correction described in eq.~(\ref{eq:rearrange}) is that adoption rates need not be strictly positive. This is, for instance, visible in the confidence intervals of the corrected usage rates in Fig.~\ref{fig:adoption}, where the confidence intervals for all countries are centered on zero in the period before the widespread availability of genAI coding tools.

\subsection{Code Verbosity}
Here, we explore whether human coding styles affects the accuracy of AI detection, focusing on verbosity. In case the AI detector misclassifies verbose human code as AI-generated, we evaluated whether verbosity or templatedness predict false positives among human-written functions.

To measure verbosity of Python functions, we used individual features common in the software engineering literature on readability and code style: average line length, blank-line ratio, comment ratio, docstring length, and token count. These capture layout, documentation, and size dimensions of code verbosity \cite{buse2009learning,miara1983program}. We construct two composite measures. The first, Composite Verbosity, averages standardized values of line length, blank-line ratio, comment ratio, and docstring length, capturing stylistic and documentation verbosity. The second, Composite Verbosity + Size, adds standardized token count to these components to incorporate overall code length alongside style and documentation.
Templatedness, representing repetitiveness or boilerplate structure, was measured as one minus normalized token entropy, following the ``naturalness of software'' literature 
\cite{hindle2016naturalness}.

\begin{table}[h!]
\centering
\begin{tabular}{lccc}
\hline
\textbf{Variable} & \textbf{Spearman $\rho$} & \textbf{t statistic} & \textbf{p-value} \\
\hline
Average line length & -0.017 & -0.526 & 0.599 \\
Blank ratio & -0.017 & -0.527 & 0.599 \\
Comment ratio & -0.006 & -0.186 & 0.852 \\
Docstring length & -0.141 & -4.515 & $<0.001$ \\
\# of Tokens & 0.109 & 3.452 & $<0.001$ \\
Composite Verbosity & -0.042 & -1.326 & 0.185 \\
Composite Verbosity + Size & -0.010 & -0.309 & 0.757 \\
Templatedness & 0.051 & 1.619 & 0.106 \\
\hline
\end{tabular}
\caption{Spearman correlations and associated test statistics measuring the relationship between measures of code verbosity of human written code and the false classification of the code as AI-written.}
\label{tab:verbosity_correlations}
\end{table}

We report two analyses on 1,000 randomly chosen out of sample Python functions written by humans. We analyze the likelihood that these functions are incorrectly classified as written by AI (false positives/FP).

\paragraph{Correlations with classifier false-positives (human code classified as AI).} Across 1,000 human functions, none of the individual verbosity features correlate strongly with FP status. The largest absolute Spearman correlation is -0.14 for docstring length (fewer FPs for more documented code). Token count shows a mild positive correlation ($\rho = 0.11$), and templatedness is weakly positively correlated ($\rho = 0.05$). The composite verbosity indices are near zero (composite verbosity: $\rho= -0.04$; composite verbosity + size: $\rho= -0.01$). Overall, verbosity does not systematically relate to the false-positive rate. We report the correlations and p-values in Table~\ref{tab:verbosity_correlations}.

\begin{figure}[h!]
    \centering
    \includegraphics[width = 0.9\linewidth]{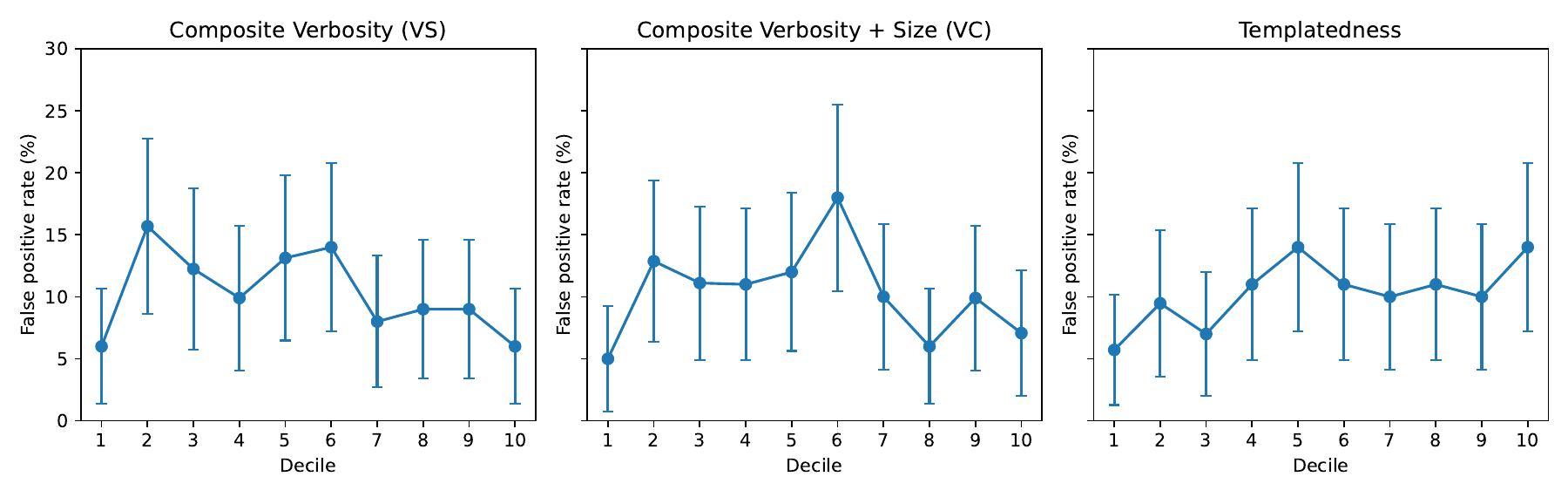}
    \caption{The relationship between code verbosity, operationalized in three ways, and the likelihood a human-written function is incorrectly classified as AI-written. We split the verbosity scores into deciles, and plot the mean false positive rate with bootstrapped confidence intervals. We find no consistent relationship between code verbosity and false positives.}
    \label{fig:verbosity_decile}
\end{figure}

\paragraph{False positives by verbosity decile.} When human functions are grouped by verbosity, FP rates fluctuate around 10\% with no monotonic trend, see Fig.\ref{fig:verbosity_decile}. This pattern suggests random variation rather than systematic bias in code verbosity among human-written functions incorrectly classified as AI written.

\subsection{Value produced by genAI coding assistants}
In the discussion section, we provide back-of-the-envelope calculations for the monetary value of genAI in US coding activities. These calculations start from our estimate that, by the end of 2024, 28.6\% of functions from US developers were produced by genAI. Note that we report the estimate as 29\% in the main text but carry out all calculations with the more precise estimate of 28.6\%. Based on our baseline, conservative point estimate of $0.122$ for the effect of genAI on commit volumes, a 29\% usage rate will translate into an overall increase of $e^{0.122*0.286}-1 =3.55\% \approx 3.6\%$. We then analyze detailed task surveys and wage statistics for about 900 different occupations. This yields that the US spends between an estimated 637 and 1,063B USD on coding-related labor costs per year (SI, section \ref{sec:SI_wagesum}). 

If we are willing to assume that (1) our estimated productivity effects (based on open-source Python contributions) are representative of code in general and (2) that the associated increase in commits accurately reflects the underlying causal effect on labor productivity of coders --- including that more commits translate into more valuable code --- we arrive at a total value of productivity increases in US coding work of between 23 and 38B USD a year. If, instead, productivity effects are negligible outside Python, the value of genAI could be as low as 17\% \cite{github_language_stats}, of our most conservative estimate above (i.e., 3.2-5.3B USD). However, not only do we deem this implausible, but absent additional information, it would also go against the maximum-ignorance stance that we implicitly took above.   

Despite controlling for observed and unobserved user characteristics, our estimation design is not optimized for identifying causal effects. Moreover, we cannot account for any spillover effects from genAI usage across workers. In fact, our estimates almost certainly understate the full economic impact of genAI in software development. Field experiments that track developers in live codebases report increases of roughly 13.6\% in commits and a 26\% faster task completion rate \cite{cuiEffectsGenerativeAI2024}. Natural experiments exploiting the staggered roll-out of GitHub Copilot find that access to Copilot leads to a 6\% increase in merged pull-requests \cite{song2024impact} and a 5.5\% increase in commits \cite{hoffmann2024generative}. In lab experiments, subjects with access to genAI complete software tasks 21-55\% faster, translating to a 6.0\%-15.7\% effect assuming a 28.6\% AI use rate \cite{peng2023impact,paradis2025aiimpact}. Combining the average of the three estimates of improvement in task-completion time from RCTs (55\% \cite{peng2023impact}, 21\% \cite{paradis2025aiimpact}, and 26\% \cite{cuiEffectsGenerativeAI2024}) with our estimated 28.6\% AI use rate would imply a 9.7\% effect and an annual value of \$62B-\$103B USD. Relying instead on the average increased commit and merged pull request rates across one RCT and two natural experiments (13.6\% \cite{cuiEffectsGenerativeAI2024}, 5.5\% \cite{hoffmann2024generative}, and 5.6\% \cite{song2024impact}) with the estimated AI use rate yields an 2.35\% increase in productivity or 15B-25B USD in annual value.

Note, however, that none of these estimates account for general equilibrium effects. If genAI raises worker output, the price of code will drop, leading to savings for its consumers. This in turn, increases the demand for code. In the SI, section~\ref{sec:SI_geneq}, we show that, depending on the exact assumptions about elasticities and market conditions one is willing to make, total welfare effects may lie anywhere between 21B and 126B USD based on our own effect estimates, with even larger gains possible if effects are closer to the ones reported in lab and field experiments. However, the upper bounds of these estimates refer to the extreme situation where supply of code is perfectly elastic. This unrealistically assumes that the drop in price now benefits consumers across the entire volume of code they consume, but does not at all  hurt producers of this code. The (equally unlikely) lower bounds, in contrast, are very close to our initial estimates that do not adjust for general equilibrium effects. These initial estimates can therefore be considered conservative, also in the longer run.

\section*{Supplementary Information}

\section{AI Detector Model Validation}

\subsection{Validation on real-world AI-generated code}

To evaluate how well our detector recognizes the use of genAI in real-world code, we rely on  conversations with LLMs recorded in the WildChat database~\cite{zhao2024wildchat}.  WildChat comprises 1M real-world user-AI conversations. We focus on  conversations containing Python functions, extracting 21k functions for interactions with GPT-3.5-turbo and 12k functions for interactions with GPT-4.

\begin{figure}[H]
    \centering
    \includegraphics[width = 0.8\linewidth]{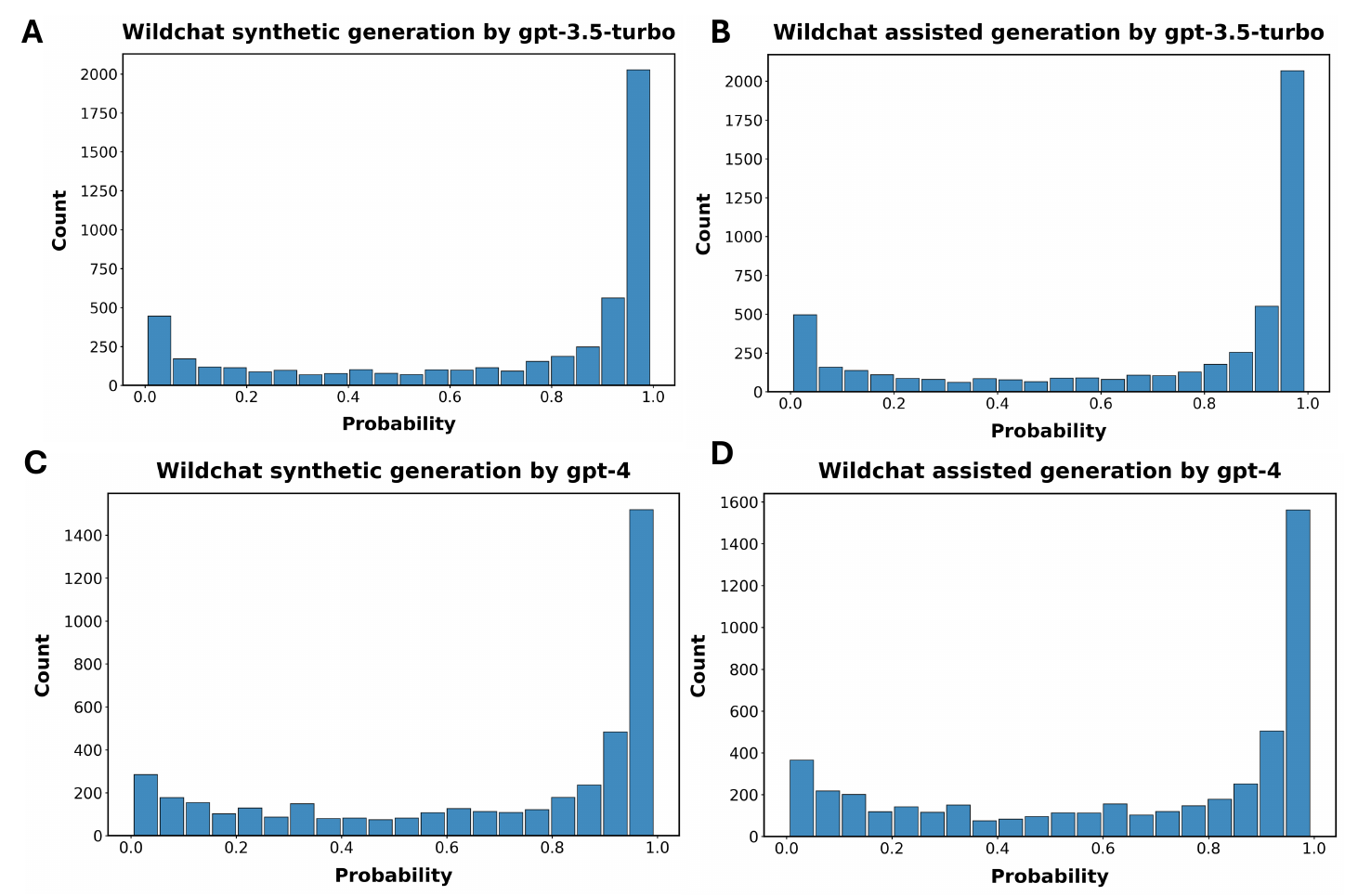}
    \caption{AI detection in functions from WildChat. The panels show histograms of the  probability that functions originally created by LLMs were detected as AI-generated. Functions are taken from interactions of human coders with GPT-3.5 turbo (panels \textbf{A} and \textbf{B}) and GPT-4 (panels \textbf{C} and \textbf{D})  collected in the WildChat dataset. ``Synthetic'' code (panels \textbf{A} and \textbf{C}) refers to code generated in the initial response of the LLM at the start of each interaction. ``Assisted code'' refers to code generated in later rounds of the interaction between the human coder and the LLM.  }
    \label{fig:performance_alt_models}
\end{figure}

Figure~\ref{fig:performance_alt_models} shows how well our detector identifies these functions as AI-generated, using a random sample of 5,000 functions for each LLM. We study functions created under two separate conditions. First, we use a sample of functions that were generated in the first round interaction with the LLMs, so called "synthetic" functions. Next, we use a sample of functions that were generated in later rounds of the interaction with the LLMs. We refer to these functions as  "assisted". The average probability of true positives is very similar  across samples at around 0.7 (GPT-3.5 turbo --- synthetic: 0.701, assisted: 0.698; GPT-4 --- synthetic: 0.675, assisted: 0.651). This suggests that although the AI detector performs somewhat less reliably on AI-generated created by real-world users, it is unaffected by \emph{how much} a user interacts with the LLM to generate the code.

\subsection{Validation on more recent genAI models}
Since our data collection ended in 2024, newer LLMs have been introduced. Here, we test the performance of our detection model on code created by four such models: OpenAI's GPT-4.1, Anthropic's Claude Sonnet 4, and Deepseek-V3. We also test the model on code generated by OpenAI's o3, a reasoning model. To do so, we reproduced the procedure depicted in Fig.~\ref{fig:1} to  generate an additional 1,000 synthetic functions for GPT-4.1 and o3, and 500 for Claude Sonnet 4 and Deepseek-V3. We then test how well the detector manages to identify these functions as AI-generated.

\begin{figure}[H]
    \centering
    \includegraphics[width = 0.7\linewidth]{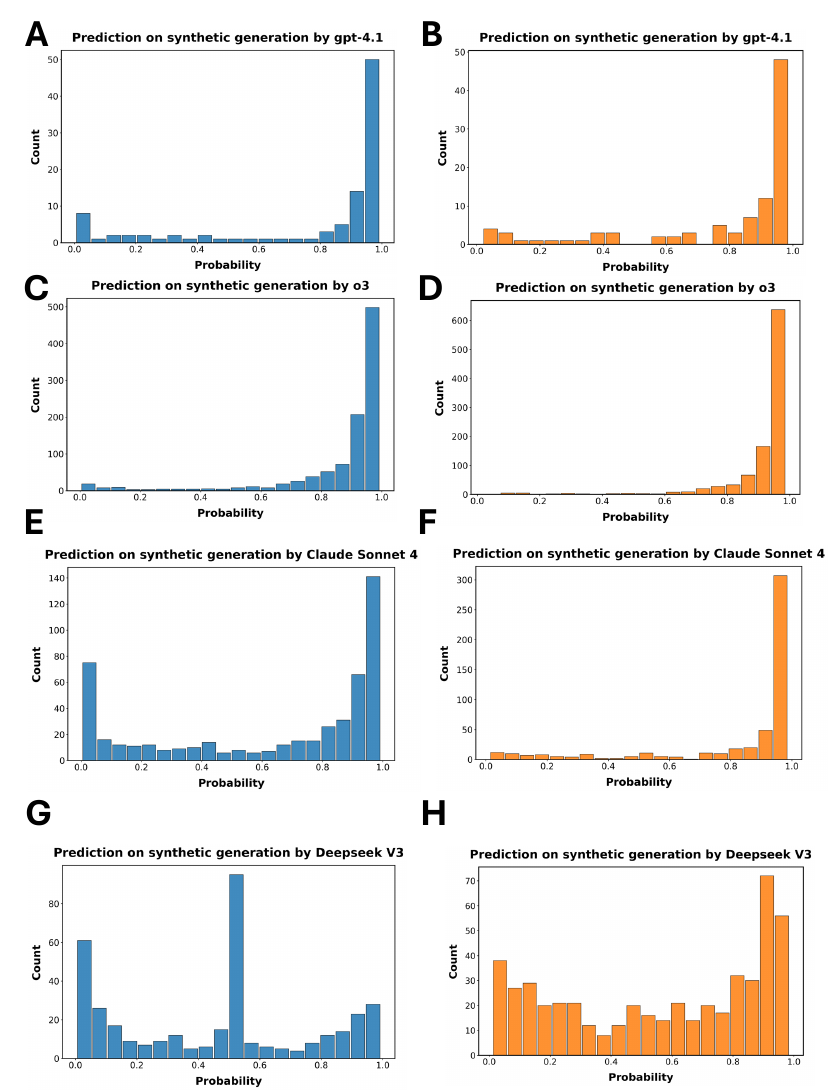}
    \caption{Performance of classifier on  code generated by recent LLMs. Each panel shows a histogram of the predicted probability that the classifier correctly identifies functions as AI-generated.  Rows correspond to the LLM used to generate the function: \textbf{A/B}: GPT-4.1; \textbf{C/D} GPT-o3; \textbf{E/F}:  Claude Sonnet 4; \textbf{G/H}: DeepSeek-V3. Panels on the left correspond to predicted probabilities without retraining the classifier; panels on the right are based on a retrained classifier that uses an additional 500 training examples of Claude- and Deepseek generated functions.}
    \label{fig:altmodels_classifier_test}
\end{figure}

Results are presented in Fig.~\ref{fig:altmodels_classifier_test}.
The detection model is still predictive, especially for the newer OpenAI models, but struggles in differentiating between human code and code generated by either Claude or DeepSeek. 
Therefore, we add 500 examples of DeepSeek- and Claude-generated functions to our training set and  re-train the model. In the case of o3, performance comes close to the 0.95 average probability of true positives we observed for our initial sample of synthetic functions. Table~\ref{tab:model-performance} shows that adding this modest number of example functions already substantially improves detection performance, even for the models that were not used in the  creation of additional training data. We therefore expect that by expanding the training set, we may be able to maintain the detector's performance also as genAI technologies improve, at least in the near future.

\begin{table}[ht]
\centering
\begin{tabular}{lcc}
\hline
\textbf{Model} & \textbf{Before Training} & \textbf{After Training} \\
\hline
\textbf{DeepSeek-V3}      & 0.452 & 0.561 \\
\textbf{Claude-Sonnet-4}  & 0.630 & 0.831 \\
\textbf{o3}               & 0.867 & 0.910 \\
\textbf{GPT-4.1}          & 0.763 & 0.781 \\
\hline
\end{tabular}
\caption{Average probability of true positives before and after training.}
\label{tab:model-performance}
\end{table}

\section{Cross-country differences}\label{sec:SI_crosscountry}

In Fig.~\ref{fig:adoption}, we show genAI adoption rates in six major countries in software development with their 95\% confidence intervals. To assess to what extent these adoption trajectories differ significantly across these countries, we first aggregate the data to the country-year level. Next, we run two-sided equality-of-means t-tests, comparing all countries to one another in each year. The results are summarized in Table~\ref{tab:pvalues_countries}. In each cell, the table lists the p-values for the comparisons in the associated country pair. p-values below 0.01 are depicted in bold. Moreover, whenever the row-country's adoption rate exceeds the column country's, the p-value is colored green. Otherwise ---when the column country leads the row country--- the p-value is colored red and put in parentheses.

\begin{table}[htbp]
\centering
\caption{Pairwise p-values for differences in AI adoption rates}
\label{tab:country_pvalues}
\scriptsize
\vspace{0.5em}% year
\begin{tabular}{lcccccc}
\hline
2019 & China & France & Germany & India & Russia & United States \\
\hline
China & -- & \textcolor{red}{(0.647)} & \textcolor{red}{(0.471)} & \textcolor{red}{(0.925)} & \textcolor{red}{(0.333)} & \textcolor{red}{(0.509)} \\
France & \textcolor{green}{0.647} & -- & \textcolor{red}{(0.965)} & \textcolor{green}{0.682} & \textcolor{red}{(0.778)} & \textcolor{green}{0.856} \\
Germany & \textcolor{green}{0.471} & \textcolor{green}{0.965} & -- & \textcolor{green}{0.505} & \textcolor{red}{(0.753)} & \textcolor{green}{0.700} \\
India & \textcolor{green}{0.925} & \textcolor{red}{(0.682)} & \textcolor{red}{(0.505)} & -- & \textcolor{red}{(0.355)} & \textcolor{red}{(0.552)} \\
Russia & \textcolor{green}{0.333} & \textcolor{green}{0.778} & \textcolor{green}{0.753} & \textcolor{green}{0.355} & -- & \textcolor{green}{0.472} \\
United States & \textcolor{green}{0.509} & \textcolor{red}{(0.856)} & \textcolor{red}{(0.700)} & \textcolor{green}{0.552} & \textcolor{red}{(0.472)} & -- \\
\hline
\end{tabular}\\
% End 
\vspace{0.5em}% y
\begin{tabular}{lcccccc}
\hline
2020 & China & France & Germany & India & Russia & United States \\
\hline
China & -- & \textcolor{green}{0.196} & \textcolor{green}{\textbf{0.000}} & \textcolor{green}{0.217} & \textcolor{green}{0.104} & \textcolor{green}{0.029} \\
France & \textcolor{red}{(0.196)} & -- & \textcolor{green}{\textbf{0.000}} & \textcolor{green}{0.935} & \textcolor{green}{0.501} & \textcolor{green}{0.192} \\
Germany & \textcolor{red}{(\textbf{0.000})} & \textcolor{red}{(\textbf{0.000})} & -- & \textcolor{red}{(\textbf{0.000})} & \textcolor{red}{(\textbf{0.000})} & \textcolor{red}{(\textbf{0.000})} \\
India & \textcolor{red}{(0.217)} & \textcolor{red}{(0.935)} & \textcolor{green}{\textbf{0.000}} & -- & \textcolor{green}{0.592} & \textcolor{green}{0.401} \\
Russia & \textcolor{red}{(0.104)} & \textcolor{red}{(0.501)} & \textcolor{green}{\textbf{0.000}} & \textcolor{red}{(0.592)} & -- & \textcolor{green}{0.987} \\
United States & \textcolor{red}{(0.029)} & \textcolor{red}{(0.192)} & \textcolor{green}{\textbf{0.000}} & \textcolor{red}{(0.401)} & \textcolor{red}{(0.987)} & -- \\
\hline
\end{tabular}\\
% End of
\vspace{0.5em}% year
\begin{tabular}{lcccccc}
\hline
2021 & China & France & Germany & India & Russia & United States \\
\hline
China & -- & \textcolor{red}{(0.395)} & \textcolor{red}{(0.261)} & \textcolor{red}{(0.170)} & \textcolor{red}{(0.533)} & \textcolor{red}{(0.088)} \\
France & \textcolor{green}{0.395} & -- & \textcolor{red}{(0.919)} & \textcolor{red}{(0.808)} & \textcolor{green}{0.764} & \textcolor{red}{(0.839)} \\
Germany & \textcolor{green}{0.261} & \textcolor{green}{0.919} & -- & \textcolor{red}{(0.870)} & \textcolor{green}{0.640} & \textcolor{red}{(0.919)} \\
India & \textcolor{green}{0.170} & \textcolor{green}{0.808} & \textcolor{green}{0.870} & -- & \textcolor{green}{0.511} & \textcolor{green}{0.894} \\
Russia & \textcolor{green}{0.533} & \textcolor{red}{(0.764)} & \textcolor{red}{(0.640)} & \textcolor{red}{(0.511)} & -- & \textcolor{red}{(0.458)} \\
United States & \textcolor{green}{0.088} & \textcolor{green}{0.839} & \textcolor{green}{0.919} & \textcolor{red}{(0.894)} & \textcolor{green}{0.458} & -- \\
\hline
\end{tabular}\\
% End o
\vspace{0.5em}% year 
\begin{tabular}{lcccccc}
\hline
2022 & China & France & Germany & India & Russia & United States \\
\hline
China & -- & \textcolor{green}{\textbf{0.000}} & \textcolor{green}{\textbf{0.000}} & \textcolor{green}{\textbf{0.000}} & \textcolor{green}{\textbf{0.001}} & \textcolor{red}{(\textbf{0.000})} \\
France & \textcolor{red}{(\textbf{0.000})} & -- & \textcolor{red}{(0.129)} & \textcolor{green}{0.154} & \textcolor{red}{(0.012)} & \textcolor{red}{(\textbf{0.000})} \\
Germany & \textcolor{red}{(\textbf{0.000})} & \textcolor{green}{0.129} & -- & \textcolor{green}{\textbf{0.008}} & \textcolor{red}{(0.421)} & \textcolor{red}{(\textbf{0.000})} \\
India & \textcolor{red}{(\textbf{0.000})} & \textcolor{red}{(0.154)} & \textcolor{red}{(\textbf{0.008})} & -- & \textcolor{red}{(\textbf{0.000})} & \textcolor{red}{(\textbf{0.000})} \\
Russia & \textcolor{red}{(\textbf{0.001})} & \textcolor{green}{0.012} & \textcolor{green}{0.421} & \textcolor{green}{\textbf{0.000}} & -- & \textcolor{red}{(\textbf{0.000})} \\
United States & \textcolor{green}{\textbf{0.000}} & \textcolor{green}{\textbf{0.000}} & \textcolor{green}{\textbf{0.000}} & \textcolor{green}{\textbf{0.000}} & \textcolor{green}{\textbf{0.000}} & -- \\
\hline
\end{tabular}\\
% End 
\vspace{0.5em}% year
\begin{tabular}{lcccccc}
\hline
2023 & China & France & Germany & India & Russia & United States \\
\hline
China & -- & \textcolor{red}{(\textbf{0.000})} & \textcolor{red}{(\textbf{0.000})} & \textcolor{red}{(0.944)} & \textcolor{red}{(\textbf{0.000})} & \textcolor{red}{(\textbf{0.000})} \\
France & \textcolor{green}{\textbf{0.000}} & -- & \textcolor{red}{(\textbf{0.001})} & \textcolor{green}{\textbf{0.001}} & \textcolor{red}{(\textbf{0.000})} & \textcolor{red}{(\textbf{0.000})} \\
Germany & \textcolor{green}{\textbf{0.000}} & \textcolor{green}{\textbf{0.001}} & -- & \textcolor{green}{\textbf{0.000}} & \textcolor{green}{0.937} & \textcolor{red}{(\textbf{0.000})} \\
India & \textcolor{green}{0.944} & \textcolor{red}{(\textbf{0.001})} & \textcolor{red}{(\textbf{0.000})} & -- & \textcolor{red}{(\textbf{0.000})} & \textcolor{red}{(\textbf{0.000})} \\
Russia & \textcolor{green}{\textbf{0.000}} & \textcolor{green}{\textbf{0.000}} & \textcolor{red}{(0.937)} & \textcolor{green}{\textbf{0.000}} & -- & \textcolor{red}{(\textbf{0.000})} \\
United States & \textcolor{green}{\textbf{0.000}} & \textcolor{green}{\textbf{0.000}} & \textcolor{green}{\textbf{0.000}} & \textcolor{green}{\textbf{0.000}} & \textcolor{green}{\textbf{0.000}} & -- \\
\hline
\end{tabular}\\
% End of year
\vspace{0.5em}% year
\begin{tabular}{lcccccc}
\hline
2024 & China & France & Germany & India & Russia & United States \\
\hline
China & -- & \textcolor{red}{(\textbf{0.000})} & \textcolor{red}{(\textbf{0.000})} & \textcolor{red}{(\textbf{0.000})} & \textcolor{red}{(\textbf{0.000})} & \textcolor{red}{(\textbf{0.000})} \\
France & \textcolor{green}{\textbf{0.000}} & -- & \textcolor{red}{(0.852)} & \textcolor{green}{\textbf{0.000}} & \textcolor{green}{\textbf{0.000}} & \textcolor{red}{(\textbf{0.000})} \\
Germany & \textcolor{green}{\textbf{0.000}} & \textcolor{green}{0.852} & -- & \textcolor{green}{\textbf{0.000}} & \textcolor{green}{\textbf{0.000}} & \textcolor{red}{(\textbf{0.000})} \\
India & \textcolor{green}{\textbf{0.000}} & \textcolor{red}{(\textbf{0.000})} & \textcolor{red}{(\textbf{0.000})} & -- & \textcolor{green}{\textbf{0.000}} & \textcolor{red}{(\textbf{0.000})} \\
Russia & \textcolor{green}{\textbf{0.000}} & \textcolor{red}{(\textbf{0.000})} & \textcolor{red}{(\textbf{0.000})} & \textcolor{red}{(\textbf{0.000})} & -- & \textcolor{red}{(\textbf{0.000})} \\
United States & \textcolor{green}{\textbf{0.000}} & \textcolor{green}{\textbf{0.000}} & \textcolor{green}{\textbf{0.000}} & \textcolor{green}{\textbf{0.000}} & \textcolor{green}{\textbf{0.000}} & -- \\
\hline
\end{tabular}\\
% End o
\caption{Comparison of mean adoption rates: t-tests. The table reports p-values of t-tests that compare the mean genAI adoption rate of the row country to the mean adoption rate of the column country in each year between 2019 and 2024. When these differences are statistically significant at the 1\% level, p-values are bold. Green colors refer to years in which the row country leads the column country in genAI adoption, red colors the opposite. In the latter case, p-values are put in parentheses.}
\label{tab:pvalues_countries}
\end{table}

\section{Detailed regression results}\label{sec:SI_regression}

To estimate the effect of AI adoption on users' coding output as well as associations between AI adoption rates and demographic variables, we first need to determine the adoption rate of a given user at a given point in time.  We focus on US users and drop those users with implausible commit frequencies, either pushing over 10k commits in total or over 2k commits in a single quarter, as likely bots or automated accounts. Next, we calculate the average score from our AI-detection algorithm --- corrected by applying applying eq.~(\ref{eq:rearrange}) --- across all functions produced by a user in each given quarter. This yields a user-quarter level dataset that describes AI usage rates of 100,097 users over a time span of, on average, 5.1 quarters. Whenever these averages are based on fewer than 10 functions, the observations are set to missing. In principle, it is also possible to reduce the number of missing observations by interpolation between non-missing observations. However, this risks using information from future quarters. To avoid this, we instead fill values forward for at most 2 quarters. That is,  we assume that --- absent further information --- AI adoption rates remain constant, at least for a limited amount of time.

\subsection{AI adoption and output}
To determine the effects of AI adoption on the quantity and nature of output that GitHub users produce, we add to our user-quarter level dataset information on the volume and types of commits that users produce, as well as the libraries they use in these commits. We analyze the effect of genAI adoption on these variables using fixed effects models. That is, we estimate Ordinary Least Squared models (OLS) with user and quarter fixed effects, $\rho_u$ and $\tau_q$. This allows us to estimate the effects of genAI by comparing how output changes with changes in AI adoption within the same user, keeping constant user characteristics that do not change over time, while also controlling for secular changes that affect all users in a given quarter. 

Our main variable of interest, ${AI}_{uq}$, measures the estimated share of functions user $u$ produced by genAI in quarter $q$, that is the average prediction of our genAI classifier for functions in this quarter. Because detection of genAI requires that we observe functions --- which in turn requires commit activity --- we measure AI adoption rates in the quarter before the commits are counted. In other words, our primary specification uses a one-quarter lagged measure of AI adoption. This avoids mechanical relations between observed AI adoption and commit-based variables. That is, we estimate:

\begin{equation}\label{eq:SI_regmodel}
    y_{uq} = \beta_{AI} {AI}_{uq-1} + \rho_u + \tau_q + \varepsilon_{uq}
\end{equation}
where $y_{uq}$ is one of our dependent variables. Finally, to determine the precision of our estimates, we use standard errors clustered by individual user, allowing for correlations in errors within individuals. 

We estimate these models for two types of dependent variables. The first type measures the activity rates of users in terms of the number of commits they push in each quarter. We create three types of commit counts. The first, $N_{uq}^\texttt{all}$, counts all commits by a user in a quarter. The second, $N_{uq}^\texttt{mult}$ counts commits that make changes to multiple files (scripts) in a project. For these commits, it is more likely that they need to navigate project-level dependencies, making them in principle more complex. The third, $N_{uq}^\texttt{imp}$, counts commits that add library imports to a script. Libraries are open-source software modules written (often) by other developers, which users ``import'' in their own scripts to use. These libraries are thought of as building blocks of modern software development \cite{eghbal2016roads}. Presumably, these commits are more likely to change major functionality in a script and are more substantial.     

The second way we study user activity aims to quantify changes in the nature of the code users produce by tracking the kind of software libraries they import in their scripts. The rationale for this is that different libraries facilitate different types of functionality, revealing information on the broad programming domain to which a script belongs. Libraries, and especially combinations of libraries used by programmers, therefore provide a rough indication of what kind of program a user works on. If we observe that users start using new libraries or library combinations, we interpret this as a sign that the user experiments with new types of code. This is in line with prior work that interprets new library combinations in scripts as a sign of innovation \cite{fang2024novelty}. 

We study two broad kinds of library usage in user commit histories.
\begin{itemize}
    \item $L_{uq}^\texttt{all}$: the number of unique libraries that a user adds across all their commits in a given quarter and
    \item $L_{uq}^\texttt{entry}$: the number of unique libraries that a user adds in quarter $q$ that had not been used in any prior quarter by the same user.
\end{itemize}  
To initialize the latter, we drop the first year of observations for each user to assess library usage that is new-to-the-user. We also generate $C_{uq}^\texttt{all}$ and  $C_{uq}^\texttt{entry}$, which count commit-level library combinations (i.e., the sets of libraries added in single commit) rather than individual library use or entry by a user. Finally, we create variables, $P_{uq}^\texttt{all}$ and  $P_{uq}^\texttt{entry}$, which count the use of, and entry into, new library pairs, any pairwise combination of libraries in the commit-level library sets used to generate the $C_{uq}$ variables.

Finally, to check the robustness of these results, we test two further changes in how we construct these library-based variables. First, if genAI adds libraries that are atypical and rarely used (so-called ``AI slop''), this may affect the analysis of new-to-the-user entries. To mitigate against this, we rerun all analyses using only the 5,000 most commonly used libraries. Second, not all libraries will radically change the domain of code on which a user works. Therefore, we also test what happens if we use the aggregate, coarsened library categories instead of the libraries themselves, described in Section \ref{sec:SI_libcommunities}. This yields the following set of dependent variables: 
\begin{itemize}
    \item 5,000 most common libraries only:
    \begin{itemize}
        \item libraries: $L_{uq}^\texttt{all,5k}$, $L_{uq}^\texttt{entry,5k}$; 
        \item library combinations: $C_{uq}^\texttt{all,5k}$, $C_{uq}^\texttt{entry,5k}$;
        \item library pairs: $P_{uq}^\texttt{all,5k}$, $P_{uq}^\texttt{entry,5k}$ 
    \end{itemize} 
    \item Coarsened library categories: 
    \begin{itemize}
        \item categories: $L_{uq}^\texttt{all,cat}$, $L_{uq}^\texttt{entry,cat}$; 
        \item category combinations: $C_{uq}^\texttt{all,cat}$, $C_{uq}^\texttt{entry,cat}$;
        \item category pairs: $P_{uq}^\texttt{all,cat}$, $P_{uq}^\texttt{entry,cat}.$
        \end{itemize}
 
\end{itemize}

We then proceed by log-transforming each of these dependent variables. To avoid $\log(0)$ issues, we increment each count by 1 unit before taking logs. We then estimate the following models:
\begin{align*}
    \log (y^\texttt{type}_{iq}+1) = \beta_{AI}^\texttt{type} {AI}_{iq-1}+ \rho_i + \tau_q + \varepsilon_{iq},
\end{align*}
where $y^\texttt{type}_{iq}$ stands for one of the various counts defined above for user $i$ in quarter $q$. $\rho_i$ denotes user fixed effects, $\tau_q$  quarter fixed effects. ${AI}_{iq}$ is the average estimated AI usage rate across commits by individual $i$ in quarter $q$, as defined in  eq.~(\ref{eq:rearrange}). As before, we lag this variable to avoid that the commit-based dependent variables are by construction related to our variable of interest. The resulting parameter estimates $\hat{\beta}_{AI}^\texttt{type}$ can be interpreted as semi-elasticities. That is they approximately describe by which percentage $y_{iq}^\texttt{type}$ changes when adoption rates go from 0 to 100\%. The results are presented as a figure in the article, and here, as a regression table: Table~\ref{tab:main_regressions_table}.
\\

\renewcommand\cellalign{t}
\begin{table}
\resizebox{\linewidth}{!}{%
\begin{tabular}{lcccccccccccc}
 & \multicolumn{3}{c}{Commits} & \multicolumn{4}{c}{Library Use} & \multicolumn{4}{c}{Library Entry} \\
\cmidrule(lr){2-4} \cmidrule(lr){5-8} \cmidrule(lr){9-12} 
& All & Multi-file & Imports & Combos & Combos-5k & Combos-cat & Libs. & Combos & Combos-5k & Combos-cat & Libs. \\
& (log+1) & (log+1) & (log+1) & (log+1) & (log+1) & (log+1) & (log+1) & (log+1) & (log+1) & (log+1) & (log+1) \\

& (1) & (2) & (3) & (4) & (5) & (6) & (7) & (8) & (9) & (10) & (11) \\

\midrule
\addlinespace
AI Use & \makecell{0.122* \\ (0.048)} & \makecell{0.071* \\ (0.032)} & \makecell{0.074* \\ (0.032)} & \makecell{0.100* \\ (0.040)} & \makecell{0.093* \\ (0.039)} & \makecell{0.082* \\ (0.036)} & \makecell{0.118* \\ (0.049)} & \makecell{0.093* \\ (0.040)} & \makecell{0.082* \\ (0.039)} & \makecell{0.057 \\ (0.033)} & \makecell{0.045 \\ (0.041)} \\
\midrule
\addlinespace
User FE & x & x & x & x & x & x & x & x & x & x & x \\
Quarter FE & x & x & x & x & x & x & x & x & x & x & x \\
\midrule
\addlinespace
Obs. & 123,428 & 123,428 & 123,428 & 123,428 & 123,428 & 123,428 & 123,428 & 123,428 & 123,428 & 123,428 & 123,428 \\
S.E. Cluster & User & User & User & User & User & User & User & User & User & User & User \\
$R^2$ & 0.636 & 0.621 & 0.610 & 0.606 & 0.604 & 0.595 & 0.586 & 0.596 & 0.593 & 0.534 & 0.497 \\
\bottomrule
\end{tabular}
}
\caption{Estimated effects of AI usage rate on commit activity, library usage, and library entry. Stars denote significance levels: *** $p<0.001$, ** $p<0.01$, * $p<0.05$. Standard errors clustered by user are reported in the parentheses.}
\label{tab:main_regressions_table}
\end{table}

\subsection{Placebo tests}\label{sec:SI_placebo}

One potential concern is that our estimates are confounded by an omitted variable. For instance, if our AI detection model systematically miss-classifies code by users that differ in their coding output, this would confound our estimates. To analyze this, we perform a placebo test: we estimate the effect of user's ---detected--- AI usage rates before the introduction of genAI tools. That is, we rerun our analyzes in a sample that only contains quarters before the year in which co-pilot was launched, i.e., before 2022. In this period, the detected AI usage shares do not carry any information on how much users rely on genAI. However, the imprecisely estimated point estimates show that the statistical power of these placebo tests is not very strong.

Results are reported in Tab.~\ref{tab:placebo_regression_table}.
Unlike our results for the entire period, we now find no evidence that (erroneously) detected AI usage is statistically significantly associated with higher levels of activity or experimentation: p-values of our AI use range from 0.11 to 0.93. The panels of Fig.~\ref{fig:verbosity_decile} similarly shows that there are no salient differences across experience categories in measured AI use before the introduction of genAI.

\renewcommand\cellalign{t}
\begin{table}
\resizebox{\linewidth}{!}{%
\begin{tabular}{lcccccccccccc}
 & \multicolumn{3}{c}{Commits} & \multicolumn{4}{c}{Library Use} & \multicolumn{4}{c}{Library Entry} \\
\cmidrule(lr){2-4} \cmidrule(lr){5-8} \cmidrule(lr){9-12} 
& All & Multi-file & Imports & Combos & Combos-5k & Combos-cat & Libs. & Combos & Combos-5k & Combos-cat & Libs. \\
& (log+1) & (log+1) & (log+1) & (log+1) & (log+1) & (log+1) & (log+1) & (log+1) & (log+1) & (log+1) & (log+1) \\

& (1) & (2) & (3) & (4) & (5) & (6) & (7) & (8) & (9) & (10) & (11) \\

\midrule
\addlinespace
AI Use & \makecell{0.136 \\ (0.085)} & \makecell{0.008 \\ (0.050)} & \makecell{0.004 \\ (0.051)} & \makecell{0.030 \\ (0.064)} & \makecell{0.027 \\ (0.063)} & \makecell{0.015 \\ (0.058)} & \makecell{0.015 \\ (0.080)} & \makecell{0.027 \\ (0.064)} & \makecell{0.017 \\ (0.063)} & \makecell{-0.005 \\ (0.055)} & \makecell{-0.052 \\ (0.070)} \\
\midrule
\addlinespace
User FE & x & x & x & x & x & x & x & x & x & x & x \\
Quarter FE & x & x & x & x & x & x & x & x & x & x & x \\
\midrule
\addlinespace
Obs. & 123,428 & 123,428 & 123,428 & 123,428 & 123,428 & 123,428 & 123,428 & 123,428 & 123,428 & 123,428 & 123,428 \\
S.E. type & User& User& User& User& User& User& User& User& User& User& User\\
$R^2$ & 0.703 & 0.681 & 0.675 & 0.666 & 0.664 & 0.651 & 0.647 & 0.657 & 0.653 & 0.600 & 0.570 \\
\bottomrule
\end{tabular}
}
\caption{Placebo tests of main regressions. Subsetting our dataset to pre-2022 activity where we expect no AI use, we find no significant relationships between detected AI use and developer behavior. Stars denote significance levels: *** $p<0.001$, ** $p<0.01$, * $p<0.05$. Standard errors clustered by user are reported in the parentheses.}
\label{tab:placebo_regression_table}
\end{table}

\subsection{Heterogeneous effects by user experience}

To test whether the effects of AI adoption differ across programmer experience levels, we extend the baseline specification by including an interaction term between AI usage and experience. We proxy for experience using a developer's tenure on GitHub, measured as the number of years since they registered. We classify users with 6 or more years of activity on GitHub as experienced programmers, and those with less than 6 years as less experienced.

We estimate the following model:
\begin{equation}
y_{uq} = \beta_{AI} AI_{uq-1} + \beta_{exp} \mathds{1}(\text{Experience}_u \geq 6) + \beta_{AI \times exp} \left( AI_{uq-1} \times \mathds{1}(\text{Experience}_u \geq 6) \right) + \rho_u + \tau_q + \varepsilon_{uq},
\end{equation}
where $\mathds{1}(\text{Experience}_u \geq 6)$ is an indicator variable equal to 1 if user $u$ has 6 or more years of activity on GitHub. The coefficient $\beta_{AI}$ captures the effect of AI adoption for less experienced users (the reference category), while $\beta_{AI} + \beta_{AI \times exp}$ represents the total effect for experienced users. The interaction coefficient $\beta_{AI \times exp}$ therefore directly tests whether the productivity and experimentation effects of genAI differ significantly between experience groups. As before, we estimate these models using OLS with user and quarter fixed effects, clustering standard errors at the user level. Results are presented in Tab.~\ref{tab:experience_interaction_table} and visualized in Figure 3D of the main text.

\renewcommand\cellalign{t}
\begin{table}
\resizebox{\linewidth}{!}{%
\begin{tabular}{lcccccccccccc}
 & \multicolumn{3}{c}{Commits} & \multicolumn{4}{c}{Library Use} & \multicolumn{4}{c}{Library Entry} \\
\cmidrule(lr){2-4} \cmidrule(lr){5-8} \cmidrule(lr){9-12} 
& All & Multi-file & Imports & Combos & Combos-5k & Combos-cat & Libs. & Combos & Combos-5k & Combos-cat & Libs. \\
& (log+1) & (log+1) & (log+1) & (log+1) & (log+1) & (log+1) & (log+1) & (log+1) & (log+1) & (log+1) & (log+1) \\
& (1) & (2) & (3) & (4) & (5) & (6) & (7) & (8) & (9) & (10) & (11) \\
\midrule
\addlinespace
High Experience & \makecell{-0.027 \\ (0.044)} & \makecell{-0.028 \\ (0.028)} & \makecell{-0.016 \\ (0.029)} & \makecell{-0.030 \\ (0.036)} & \makecell{-0.029 \\ (0.035)} & \makecell{-0.022 \\ (0.032)} & \makecell{-0.024 \\ (0.043)} & \makecell{-0.035 \\ (0.036)} & \makecell{-0.035 \\ (0.035)} & \makecell{-0.023 \\ (0.030)} & \makecell{-0.026 \\ (0.036)} \\
AI Use & \makecell{0.020 \\ (0.064)} & \makecell{-0.030 \\ (0.040)} & \makecell{-0.018 \\ (0.041)} & \makecell{-0.001 \\ (0.051)} & \makecell{-0.004 \\ (0.051)} & \makecell{-0.008 \\ (0.046)} & \makecell{0.004 \\ (0.064)} & \makecell{-0.013 \\ (0.051)} & \makecell{-0.018 \\ (0.050)} & \makecell{-0.034 \\ (0.043)} & \makecell{-0.052 \\ (0.055)} \\
\makecell{High Experience\\ $\times$ AI Use}& \makecell{0.189* \\ (0.086)} & \makecell{0.188*** \\ (0.055)} & \makecell{0.171** \\ (0.056)} & \makecell{0.187** \\ (0.071)} & \makecell{0.179** \\ (0.069)} & \makecell{0.166** \\ (0.063)} & \makecell{0.211* \\ (0.086)} & \makecell{0.196** \\ (0.069)} & \makecell{0.186** \\ (0.068)} & \makecell{0.169** \\ (0.059)} & \makecell{0.179* \\ (0.073)} \\
\midrule
\addlinespace
User FE & x & x & x & x & x & x & x & x & x & x & x \\
Quarter FE & x & x & x & x & x & x & x & x & x & x & x \\
\midrule
\addlinespace
Observations & 123,428 & 123,428 & 123,428 & 123,428 & 123,428 & 123,428 & 123,428 & 123,428 & 123,428 & 123,428 & 123,428 \\
S.E. type & User& User& User& User& User& User& User& User& User& User& User\\
$R^2$ & 0.636 & 0.621 & 0.610 & 0.606 & 0.605 & 0.595 & 0.586 & 0.597 & 0.593 & 0.534 & 0.497 \\
\bottomrule
\end{tabular}
}
\caption{Estimated effects of AI usage rate by high vs low experience users on commit activity, library usage, and library entry. Stars denote significance levels: *** $p<0.001$, ** $p<0.01$, * $p<0.05$. Standard errors clustered by user are reported in the parentheses.}
\label{tab:experience_interaction_table}
\end{table}

\subsection{AI adoption and user demographics} 
Next, we estimate how AI adoption rates themselves differ by experience and gender, focusing on US-based users. We proxy experience as the number of years since a user's first recorded activity on GitHub. Because our data starts in 2011, this experience is right-censored. The longest experience category therefore contains individuals with the stated number of years of experience or more. Next, we regress adoption rates in 2024 on a set of dummies that encode the user's years of experience. We calculate robust (HC1) standard errors. Point estimates with their confidence intervals are plotted in Fig.~\ref{fig:3}, further details are provided in Table~\ref{tab:ai_experience}. 

\begin{table}[htbp]
\centering
\renewcommand\cellalign{t}
\renewcommand\cellgape{\Gape[4pt]}
\renewcommand{\arraystretch}{1.1}
\begin{tabular}{ll ll}
\toprule
\multicolumn{2}{c}{\textbf{Panel}} & \multicolumn{2}{c}{\textbf{Panel}} \\
\cmidrule(lr){1-2} \cmidrule(lr){3-4}
Exp 2  & \makecell{-0.017 \\ (0.013)} & Exp 10 & \makecell{-0.075*** \\ (0.013)} \\
Exp 4  & \makecell{-0.026* \\ (0.013)} & Exp 12 & \makecell{-0.083*** \\ (0.014)} \\
Exp 6  & \makecell{-0.052*** \\ (0.013)} & Exp 14 & \makecell{-0.099*** \\ (0.014)} \\
Exp 8  & \makecell{-0.066*** \\ (0.013)} & Intercept & \makecell{0.371*** \\ (0.012)} \\
\midrule
\multicolumn{4}{l}{Observations: 27,369 \quad \quad $R^2$: 0.008 \quad \quad S.E. Clustered by: \texttt{User}} \\
\bottomrule
\end{tabular}
\caption{Effect of Experience on AI Share. Standard errors clustered at the user level. Significance levels: *** $p<$0.001, ** $p<$0.01, * $p<$0.05.}
\label{tab:ai_experience}
\end{table}

\subsubsection*{Gender}

To estimate usage rates by gender, we first infer a user's gender using Gender-Guesser \url{https://pypi.org/project/gender-guesser/}, a dictionary-based method to infer a user's gender based on their first name and country (here, the US). As a consequence, this gender may not accurately reflect the gender the user identifies with. The accuracy of name-based gender inference is known to vary across nationalities \cite{santamaria2018comparison}, though this is limited to some extent by our focus on US-based developers and in general the accuracy of these methods is high when checked against ground-truth (95\%+) \cite{vanhelene2024inferring}.

Gender-Guesser returns five gender categories (``male'', ``mostly\_male'', ``andy''[androgenous], ``mostly\_female'' and ``female''), indicating the statistical confidence of the classification. Names outside the method's training data or with too few examples are classified as unknown. Applying the method to the US-based developers active in 2024, we find a male-dominated population (see Tab.~\ref{tab:gender_descriptives}). The ratio of approximately 10:1 men to women on GitHub aligns closely with estimates from previous work on gender and participation in OSS using name-based inference \cite{vasilescu2015gender} and self-reported gender identity \cite{terrell2017gender}.

\begin{table}[h!]
\centering
\begin{tabular}{lr}
\hline
\textbf{Inferred Gender} & \textbf{Count} \\
\hline
``unknown''        & 18,809 \\
``male''           & 9,531  \\
``mostly\_male''   & 1,001  \\
``female''         & 983   \\
``andy''           & 361   \\
``mostly\_female'' & 204   \\
\hline
\end{tabular}
\caption{Distribution of US GitHub users across inferred genders in 2024.}
\label{tab:gender_descriptives}
\end{table}

In our primary analysis, we only consider ``male'' and ``female'' identified users, mapping the other cases to ``unknown''. We carry out a similar regression analysis to the study of tenure and AI adoption, replacing the experience dummies by a gender dummy. Point estimates with their confidence intervals are plotted in Fig.~\ref{fig:3}, estimates are provided in Table~\ref{tab:ai_gender}. Repeating our analysis including the less certain gender identifications (``mostly\_male'' and ``mostly\_female'' as male and female, respectively), we find similar results.

\begin{table}[htbp]
\centering
\renewcommand\cellalign{t}
\renewcommand\cellgape{\Gape[4pt]}
\begin{tabular}{l c}
\toprule
& \makecell{AI Share (2024) \\ (1)} \\
\midrule
Male & \makecell{-0.004 \\ (0.013)} \\
Unknown & \makecell{0.009 \\ (0.013)} \\
Intercept (Female) & \makecell{0.431*** \\ (0.012)} \\
\midrule
Observations & 23,292 \\
S.E. Clustered & \makecell{Robust} \\
$R^2$ & 0.000 \\
\bottomrule
\end{tabular}
\caption{Relationship between user Gender and AI Share in 2024, US-based developers active in 2024. Robust standard errors. Significance levels: *** $p<$0.001, ** $p<$0.01, * $p<$0.05.}
\label{tab:ai_gender}
\end{table}

\begin{table}[htbp]
\centering
\renewcommand\cellalign{t}
\renewcommand\cellgape{\Gape[4pt]}
\begin{tabular}{l c}
\toprule
& \makecell{AI Share (2024) \\ (1)} \\
\midrule
Male & \makecell{-0.005 \\ (0.011)} \\
Unknown & \makecell{0.004 \\ (0.010)} \\
Intercept (Female) & \makecell{0.415*** \\ (0.010)} \\
\midrule
Observations & 32,386 \\
S.E. Type & \makecell{Robust} \\
$R^2$ & 0.000 \\
\bottomrule
\end{tabular}
\caption{Relationship between user Gender and AI Share in 2024, US-based developers active in 2024, using a less strict gender identification method. User display names classified by our gender inference tool as ``mostly male'' and ``mostly female'' are counted as male and female in this analysis; in the main analysis of the paper these are classified as unknown.  Robust standard errors. Significance levels: *** $p<$0.001, ** $p<$0.01, * $p<$0.05.}
\label{tab:ai_gender_robust}
\end{table}

\subsection{Measurement error}\label{sec:SI_measurement_error}

The main variable of interest in our study is a user's AI adoption rate, defined as the (latent) likelihood that a user uses genAI to program. To estimate this likelihood we use the average score from our AI-detection algorithm across all functions produced by a user in each given quarter. The more functions a user commits, the less noisy this estimate will be. That is, the precision of the estimate of AI adoption rates will depend on the number of functions a user produces in each quarter. However, in general, AI adoption rates will be measured with some error. 

Measurement error is known to lead to attenuation bias, which typically biases the estimated effect of a variable towards zero. To see this, we introduce the following notation:

\begin{itemize}
    \item ${AI}_{uq}^{*}$: real AI usage rate of a user $u$ in quarter $q$, with variance $\sigma_{AI^*}^2$;
\item ${AI}_{uq}={AI}_{uq}^{*}+\eta_{uq}$: observed (noisy) estimate of  ${AI}_{uq}^{*}$, with variance $\sigma_{AI}^2$; and
\item $\eta_{uq}$: error term that is independently distributed  from ${AI}_{uq}^{*}$, with mean zero and variance $\sigma_\eta^2$.

 \end{itemize}

In the paper, we estimate regression models of the form:

\begin{align}\label{eq:SI_regmodel_error}
    y_{uq} &= b {AI}_{uq-1} + \rho_u + \tau_q + \varepsilon_{uq} \\
    &= \beta {AI}_{uq-1}^* + \beta \eta_{uq-1}+ \rho_u + \tau_q + \varepsilon_{uq},
\end{align}
where $y_{uq}$ is one of the dependent variables we study and $\rho_u$ and $\tau_q$ are user and quarter fixed effects, respectively. The relation in this equation between the true effect of AI, $\beta$, and the estimated effect, $\hat{b}$, of the observable, but mismeasured variable, ${AI}_{uq}$, is (e.g., \cite{griliches1986errors}:

\begin{equation}\label{eq:SI_measurementerror}
    \hat{b} = \left( 1- \frac{\sigma_\eta^2}{\sigma_{AI}^2} \right) \beta 
\end{equation}

Note that due to the assumed statistical independence between AI adoption rates and measurement errors, the attenuation factor, $ 1- \frac{\sigma_\eta^2}{\sigma_{AI}^2} = 1- \frac{\sigma_\eta^2}{\sigma_{AI^*}^2 + \sigma_\eta^2}$, always lies between 0 and 1, such that the estimated effects are biased toward zero. Moreover, this term tends to 1 as measurement error becomes smaller. Consequently, $\sigma_\eta^2 \rightarrow 0 \Rightarrow  \hat{b} \rightarrow \beta$: the smaller the measurement error, the closer the estimated effect will be to the true effect.

Although the amount of measurement error is unknown, we can create different versions of ${AI}_{uq}$ across which measurement error varies in a known way. To do so, we create variants of ${AI}_{uq}$ that are based on moving averages. In particular, we average a user's detected genAI probability over a fixed number of functions, $k$:

\begin{equation}
    {AI}_{ut}^k= \frac{1}{k} \sum_{\theta =t-floor(k/2)}^{ t+ceil(k/2)-1} {AI}_{uf(\theta)},
\end{equation}
where ${AI}_{uf(\theta)}$ is the estimated probability that function $f(\theta)$ produced by user $u$ with temporal order $\theta$ was AI generated. If the window over which functions need to be collected stretches across more than 184 days (the maximum number of days in two consecutive quarters in our data), we risk that the underlying ${AI}_{ut}^*$ changes too much, and we drop the observation.

Next, for each user–quarter combination we find the two points $t_1$ and $t_2$  closest to the midpoint of a given quarter and linearly interpolate between ${AI}_{ut_1}^k$ and ${AI}_{ut_2}^k$ to arrive at an estimate of the user’s AI adoption rate in that quarter. To avoid interpolations across too long time periods, we drop user-quarter observation whenever two observations are over 184 days apart.
Using this procedure, we produce the following variables: ${AI}_{uq}^{4}$, ${AI}_{uq}^{8}$, ${AI}_{uq}^{16}$ and ${AI}_{uq}^{32}$. 

Assuming observations are identically and independently distributed (IID), the measurement error variance in each of these variables will equal $\frac{1}{k}\sigma^2_{\phi}$,
where $\sigma^2_{\phi}$ is the variance of the measurement error in a single function. This means that for any $k$: $\hat{b} = \left( 1- \frac{1}{k}\frac{\sigma_\phi^2}{\sigma_{AI^*}^2 + \frac{1}{k}\sigma_\phi^2} \right) \beta$. Consequently, attenuation bias should rise with $1/k$ approaching a linear relation as $k$ grows large . 

\begin{figure}
    \centering
            \includegraphics[width=\linewidth]{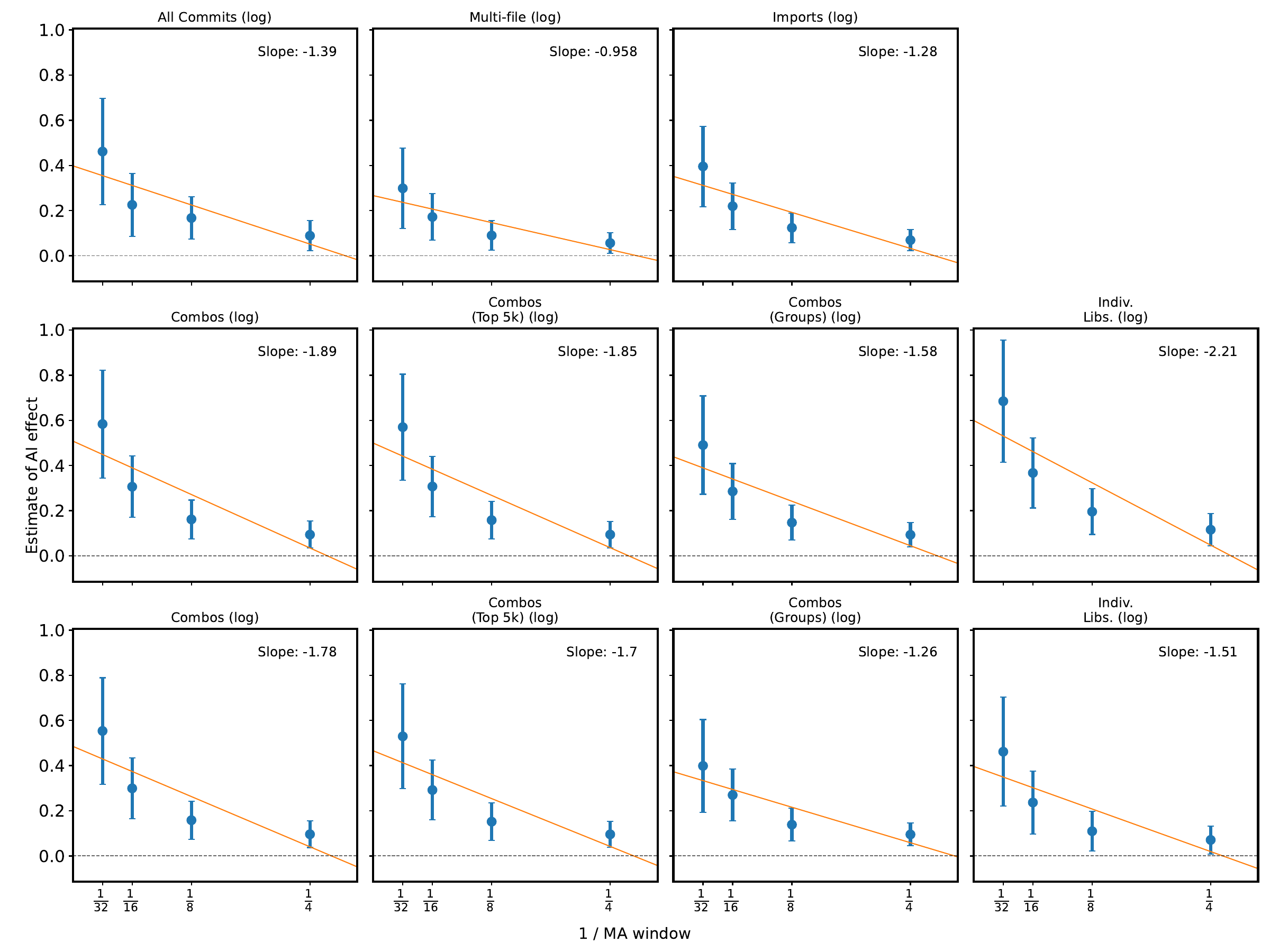}\par
    \caption{Effect estimates against measurement error. Within a specific subplot, each marker indicates the estimated effect of genAI adoption on users' coding behavior, following our baseline specification using OLS regressions with user and quarter fixed effects. The row correspond to our three main measures of user coding behavior: commits, library use, and library entry. Vertical lines indicate 95\% confidence intervals based on standard errors clustered by user.}
    \label{fig:SI_measurementerror}
\end{figure}

To test this, we rerun our baseline regression model with each of the variants ${AI}_{ut}^k$ separately, leading to four different estimates, $\hat{b}_k$. Because the number of observations for which we are able to calculate $\hat{b}_k$ drops as $k$ rises, we restrict the sample to observations where ${AI}_{ut}^k$ is nonmissing for all $k\in\{4,8,16,32\}$. In Fig.~\ref{fig:SI_measurementerror}, we plot these estimates, together with their 95\% confidence intervals, against $1/k$ for each dependent variable.

Across all models,  as measurement error falls, effect estimates rise. Measurement errors are therefore likely to lead to conservative  estimates. Moreover, at very low levels of measurement error, estimated effects on commit volumes are much closer to those reported in RCTs than our baseline estimates. 

\subsection{Nonlinearities}
In this section, we  assess whether the effect of genAI adoption on commit rates is linear or exhibits nonlinearities, such as threshold effects. To do so, we turn our genAI adoption variable into a categorical variable that collects observations in the same quintile of AI adoption. Next, we adjust the regression model of eq.~(\ref{eq:SI_regmodel}) as follows:
\begin{equation}\label{eq:SI_nonlinerarities}
    y_{uq} = \sum_{c=2}^5 \beta_{AI}^c \mathds{1}({Q_{c}^{l} \leq {AI}_{uq-1}\leq Q_{c}^{u}}) + \rho_u + \tau_q + \varepsilon_{uq}
\end{equation}
where $\mathds{1}({Q_{c}^{l} \leq {AI}_{uq-1}\leq Q_{c}^{u}})$ is a function that evaluates to 1 if a user's AI adoption falls between the lower and upper bounds ($Q_{c}^{l}$ and $Q_{c}^{u}$) of quintile $c$. The lowest quintile serves as reference category.

\renewcommand\cellalign{t}
\begin{table}
\centering
\begin{tabular}{lccc}
\toprule
 & \multicolumn{3}{c}{Commits} \\
\cmidrule(lr){2-4}
 & All Commits  & Multi-file  & Imports \\
 & (log +1) & (log +1) & (log +1) \\
 & (1) & (2) & (3) \\
\midrule
\addlinespace
AI Use 2nd Quint. & \makecell{0.065* \\ (0.033)} & \makecell{0.037 \\ (0.021)} & \makecell{0.050* \\ (0.021)} \\
AI Use 3rd Quint. & \makecell{0.109** \\ (0.034)} & \makecell{0.057** \\ (0.021)} & \makecell{0.066** \\ (0.022)} \\
AI Use 4th Quint. & \makecell{0.084* \\ (0.036)} & \makecell{0.054* \\ (0.023)} & \makecell{0.062** \\ (0.023)} \\
AI Use 5th Quint. & \makecell{0.093* \\ (0.039)} & \makecell{0.052* \\ (0.026)} & \makecell{0.054* \\ (0.026)} \\
\midrule
\addlinespace
Quarter FE & x & x & x \\
User FE & x & x & x \\
\midrule
\addlinespace
Observations & 123,428 & 123,428 & 123,428 \\
S.E. clustered on & User & User & User \\
$R^2$ & 0.636 & 0.621 & 0.610 \\
\bottomrule
\end{tabular}
\caption{Investigation of potential nonlinearity in the relationship between AI use and commit activity. The first (lowest) quintile of AI use serves as reference category. Stars denote significance levels: *** $p<0.001$, ** $p<0.01$, * $p<0.05$.}
\label{tab:ai_nonlin}
\end{table}

Results are shown in Tab.\ref{tab:ai_nonlin}. Although point estimates suggest there may be a threshold above which benefits of additional AI usage rates flatten off, the precision of these estimates is too low to warrant strong conclusions.

\section{Coarsening library information into categories}\label{sec:SI_libcommunities}

Our analysis of libraries is based on the idea that libraries provide information about the general programming domains to which code belongs (e.g., visualization, machine learning, front-end development, DevOps, etc.). A possible objection to this analysis is that genAI may add irrelevant or redundant libraries to a commit. We mitigate against this by creating coarsened categories of libraries that collect similar or related libraries. To do so, we study the co-occurrence patterns of libraries in Python projects, assuming that libraries that often co-occur in a project are likely to be related in the sense that they are either synergistic or similar in usage.

We first collect all libraries imported in our dataset of commits to Python projects. This yields over 150k libraries whose frequency roughly follows a scale-free distribution. Somewhat over a third of libraries feature only once across all projects. 

\begin{figure}[htbp]
    \centering
    \includegraphics[width=0.95\textwidth]{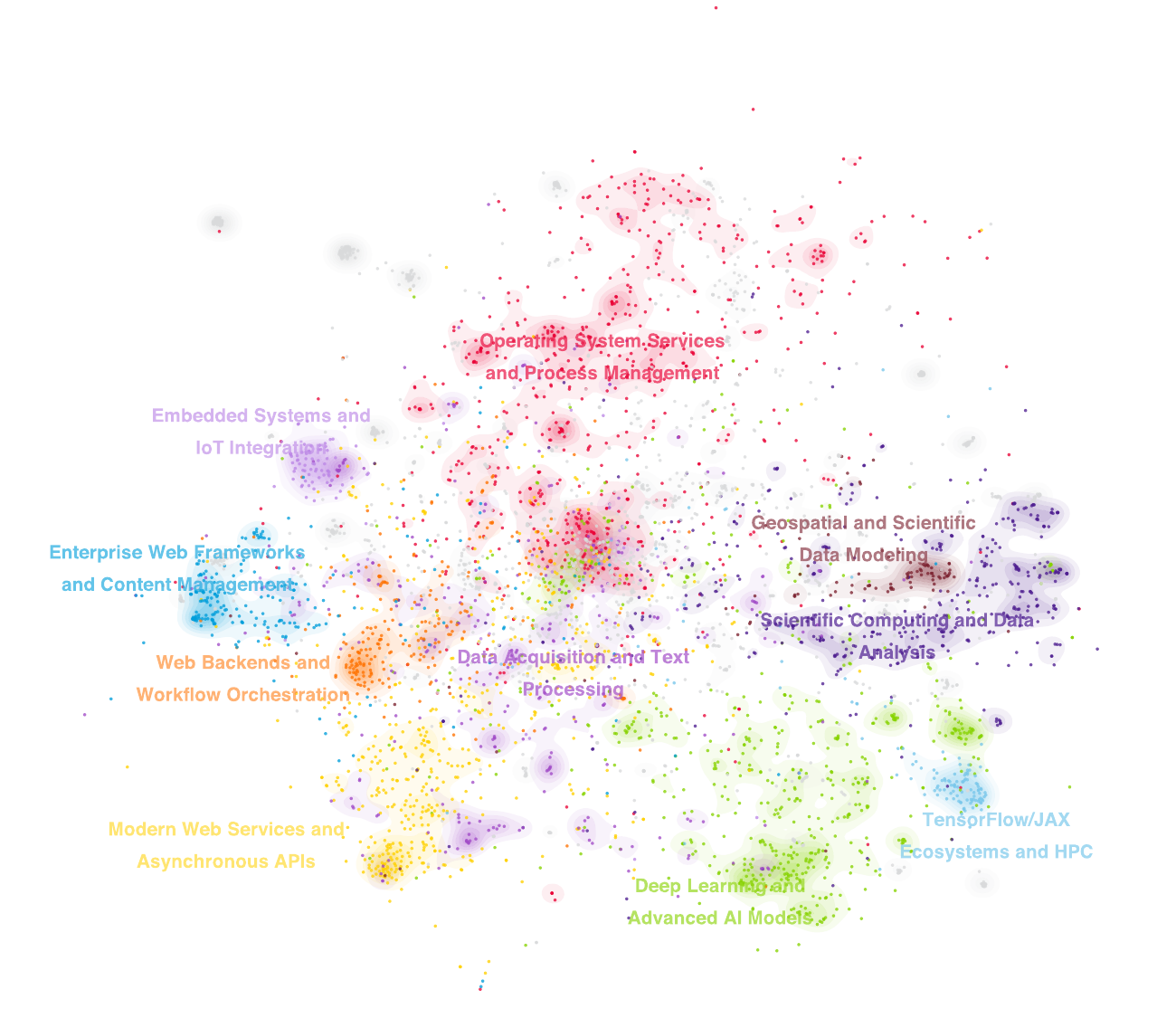}
    \caption{Library network with communities detected by a Louvain algorithm. Colors mark the top 10 of the 124 different low-level  communities, with descriptions generated by Gemini 2.5. Edges are not drawn to avoid cluttering the graph.}
    \label{fig:library_network}
\end{figure}

Next, we count how often two libraries co-occur in the same projects. To assess how surprising these co-occurrences are, we rely on an information-theoretic approach that provides Bayesian estimates of point-wise mutual information (PMI) ~\cite{van2023information}. We then construct a library network, connecting libraries that surprisingly often co-occur, using  ${PMI} > 0$ as a threshold. All pairs meeting this condition are connected by ties whose weight corresponds to the estimated PMI. We then filter the network for the 5,000 most commonly used libraries. Dropping two isolated nodes, this yields a network of 4,998 nodes connected by 186,203 edges (see Fig.~\ref{fig:library_network}).

Finally, we run a Louvain community detection algorithm to identify   library communities in this weighted graph. We obtain 124 communities, which can be further aggregated into 19 higher-level communities. From these communities, we derive alternative measures of library use, combination, and entry by users. For example, in one operationalization, we count how often a user combines a new pair of library communities in a quarter. In Table~\ref{tab:community_library}, we list representative libraries for each of these high-level communities. To help interpret these lists, we generate descriptions for each community by feeding the complete list of libraries to Gemini 2.5, using the following prompt:

\begin{promptbox}
I would like you to generate short descriptions of each of the 19 lists of Python libraries:

`` ''
Specifically, for each library lists, I want to list a domain in software development for which these libraries are typically used. Here are five examples of such domains:

1) User interfaces;
2) Data science and analytics;
3) Web design;
4) Databases;
5) Embedded systems and IoT.

Be sure to keep the descriptions short. Also, ensure that no two descriptions are too similar so that is easy to understand the differences between categories. Finally, do not concatenate two different types of domains by adding ``or'' but rather try to generalize across them or, if that is too hard, pick the most important one.

The Python libraries are separated by ','. Lists are enclosed in '[]' and the next list is always on the next line.
\end{promptbox}

\begingroup
\footnotesize
\setlength{\tabcolsep}{4pt}
\renewcommand{\arraystretch}{1.04}
\setlength{\LTpre}{0pt} \setlength{\LTpost}{6pt}
\setlength{\LTcapwidth}{\textwidth}
\begin{longtable}{@{}%
>{\raggedleft\arraybackslash}p{0.05\textwidth}%
>{\raggedleft\arraybackslash}p{0.075\textwidth}%
p{0.8\textwidth}@{}}
\toprule
\textbf{ID} & \textbf{\# libs} & \textbf{description / representative libraries} \\
\midrule
\endfirsthead
   \caption{High level library communities. Labels are generated by Gemini 2.5, based on the libraries that belong to each community. Table contains 19 high-level communities, as well as a list of the 10 most common libraries for each community. The primary statistical analyses in the paper are based on the 124 low-level communities.\label{tab:community_library}}
   \endlastfoot
\toprule
\textbf{ID} & \textbf{\# libs} & \textbf{description / representative libraries (cont.)} \\
\midrule
\endhead

\bottomrule
\endfoot

1  & 923 & \textbf{System \& Operating System Utilities:} Libraries focusing on interacting with the operating system, file system, process management, and core interpreter functions.
\\[-2pt]
   &     & \textit{\scriptsize os, sys, re, subprocess, io, tempfile, shutil, contextlib, inspect, threading} \\
\addlinespace[2pt]

2  & 903 & \textbf{Scientific Computing, AI \& Machine Learning:} Core numerical computation, deep learning frameworks (PyTorch, TensorFlow), and utilities for research and data analysis.
\\[-2pt]
   &     & \textit{\scriptsize torch, functools, argparse, math, random, tensorflow, transformers, PIL, tqdm, absl} \\
\addlinespace[2pt]

3  & 831 & \textbf{General Application Development \& Testing:} Tools for code quality, rigorous software testing (pytest, unittest), asynchronous programming, and configuration.
\\[-2pt]
   &     & \textit{\scriptsize typing, \_\_future\_\_, pytest, unittest, logging, json, collections, dataclasses, tests, homeassistant} \\
\addlinespace[2pt]

4  & 730 & \textbf{Data Science, Statistics, and Visualization:} Foundational libraries for complex data manipulation, statistical modeling, data visualization (matplotlib, seaborn), and scientific research.
\\[-2pt]
   &     & \textit{\scriptsize numpy, warnings, pandas, copy, itertools, matplotlib, scipy, ray, pickle, sklearn} \\
\addlinespace[2pt]

5  & 319 & \textbf{Web Scraping, Content \& Desktop UI:} Tools for making HTTP requests, parsing HTML, developing desktop user interfaces (tkinter), and handling multimedia/documents.
\\[-2pt]
   &     & \textit{\scriptsize requests, csv, streamlit, bs4, tkinter, sqlite3, unicodedata, html, odoo, lxml} \\
\addlinespace[2pt]

6  & 273 & \textbf{Enterprise Web \& Backend Development:} Frameworks like Django for large-scale web applications, REST APIs, time/date handling, and authentication.
\\[-2pt]
   &     & \textit{\scriptsize datetime, django, decimal, common, rest\_framework, dateutil, pytz, sentry, core, calendar} \\
\addlinespace[2pt]

7  & 262 & \textbf{Data Workflow \& Web Service Backend:} Frameworks for defining and scheduling data pipelines (Airflow), building web APIs (Flask), and interacting with various database backends (SQLAlchemy).
\\[-2pt]
   &     & \textit{\scriptsize airflow, sqlalchemy, flask, app, config, werkzeug, superset, galaxy, lib, pymongo} \\
\addlinespace[2pt]

8  & 211 & \textbf{Tooling, Packaging, and Documentation:} Utilities for code parsing (ast), dependency management (packaging), configuration file handling (toml, yaml), and generating documentation (Sphinx, Jinja2).
\\[-2pt]
   &     & \textit{\scriptsize pathlib, yaml, packaging, ast, pygments, jinja2, sphinx, tornado, docutils, mypy} \\
\addlinespace[2pt]

9  & 153 & \textbf{Networking, Asynchronous, and Blockchain:} Libraries for network programming (Twisted), cryptographic utilities, and tools related to distributed systems and blockchain technologies.
\\[-2pt]
   &     & \textit{\scriptsize mock, attr, twisted, jsonschema, synapse, web3, overrides, zope, eth\_utils, toolz} \\
\addlinespace[2pt]

10 & 122 & \textbf{Embedded Systems and IoT:} Libraries designed for running on microcontrollers and single-board computers (MicroPython, RPi), handling low-level hardware I/O.
\\[-2pt]
   &     & \textit{\scriptsize time, gc, array, board, secrets, micropython, digitalio, machine, ujson, busio} \\
\addlinespace[2pt]

11 & 92  & \textbf{Computer-Aided Design (CAD) and GUI:} Tools for geometric modeling (FreeCAD, Part), computer graphics, and building advanced, cross-platform graphical user interfaces (PyQt5).
\\[-2pt]
   &     & \textit{\scriptsize PyQt5, FreeCAD, PySide, FreeCADGui, PySide2, mantid, PySide6, PyQt6, angr, gnuradio} \\
\addlinespace[2pt]

12 & 59  & \textbf{Cloud and Infrastructure Automation:} Libraries for interacting with major cloud providers (Azure, OCI), building command-line interfaces (knack), and infrastructure testing.
\\[-2pt]
   &     & \textit{\scriptsize azure, msrest, oci, devtools\_testutils, knack, msrestazure, jmespath, litex, migen, c7n} \\
\addlinespace[2pt]

13 & 48  & \textbf{High-Energy Physics \& Scientific Simulation:} Specialized frameworks (CMSSW, ROOT) used in fields like particle physics for event simulation, data processing, and analysis.
\\[-2pt]
   &     & \textit{\scriptsize FWCore, Configuration, ROOT, Geometry, RecoTracker, DQMServices, PhysicsTools, DQM, L1Trigger, RecoMuon} \\
\addlinespace[2pt]

14 & 32  & \textbf{Scientific Data Analysis \& Crystallography:} Tools focused on processing and visualizing scientific data, particularly within crystallography, materials science (cctbx), and image analysis (dxtbx). \\[-2pt]
   &     & \textit{\scriptsize wx, libtbx, watchdog, dials, iotbx, scitbx, colors, mooseutils, cctbx, gui} \\
\addlinespace[2pt]

15 & 18  & \textbf{Natural Language Processing (NLP) \& ML Tasks:} Highly specific wrappers and scripts for common NLP and deep learning tasks like question answering, summarization, and translation.\\[-2pt]
   &     & \textit{\scriptsize tf\_keras, run\_translation, run\_qa, run\_ner, run\_mlm, run\_clm, run\_swag, run\_summarization, run\_image\_classification, run\_generation} \\
\addlinespace[2pt]

16 & 6   & \textbf{Version Control and Git Workflow:} Scripts and utilities dedicated to managing Git repositories, handling pull requests, and automating branch merging processes.
\\[-2pt]
   &     & \textit{\scriptsize gitutils, trymerge, github\_utils, label\_utils, trymerge\_explainer, test\_trymerge} \\
\addlinespace[2pt]

17 & 6   & \textbf{Large-Scale Systems Monitoring and Diagnostics:} Libraries for health checks, data collection, and diagnostic logging, often used for monitoring complex distributed systems.
\\[-2pt]
   &     & \textit{\scriptsize syscore, sysdata, systems, sysobjects, sysproduction, syslogdiag} \\
\addlinespace[2pt]

18 & 6   & \textbf{Database Testing and Benchmarking:} Tools explicitly designed for creating test harnesses, scenarios, and running rollbacks to validate the behavior of database systems (WiredTiger).
\\[-2pt]
   &     & \textit{\scriptsize wttest, wiredtiger, wtscenario, wtdataset, suite\_subprocess, test\_rollback\_to\_stable01} \\
\addlinespace[2pt]

19 & 4   & \textbf{Torch/ML Internal Development \& Testing:} Internal testing and utility modules for the PyTorch ecosystem, focusing on deep learning performance, dynamic shapes, and core framework stability.
\\[-2pt]
   &     & \textit{\scriptsize test\_torchinductor, test\_torchinductor\_dynamic\_shapes, test\_cpu\_repro, test\_aot\_inductor\_utils} \\

\bottomrule
\end{longtable}

\endgroup

\section{Estimating the total wage sum related to programming tasks in the US}\label{sec:SI_wagesum}
Estimating how much the US economy spends on programming tasks is not trivial. On the one hand, although the US Bureau of Labor Statistics' (BLS) Standard Occupation Classification (SOC) lists programming-related occupations, such as \emph{Computer programmers} and \emph{Software developers}, these jobs entail more than just coding tasks. On the other hand, workers in a wide range of other occupations may not focus on programming but still carry out substantial programming tasks, from \emph{Online merchants} to \emph{Statisticians}. To estimate how much time workers in each of the almost 900 occupations in the SOC classification spend on programming tasks, we rely on data from the Occupational Information Network (O*NET).  Next, we combine this estimate with information from the BLS' Occupational Employment and Wage Statistics (OEWS) and the American Community Survey (ACS) on employment and wages in these occupations to arrive at an estimate of the overall wage sum in the US that is associated with programming tasks. We link these data sources  at the most detailed, 6-digit, level of the SOC classification.

O*NET is the primary data source for occupational information in the US. It conducts surveys and expert analysis of occupations to determine a variety of characteristics of jobs \cite{handel2016net}. Here, we mainly use the information O*NET contains on the tasks that occupations require. We focus on the \emph{Task Ratings} file of O*NET 29.2, released in February 2025, which lists around 20 tasks for each occupation, amounting to about 17k distinct tasks. For each task, this file lists how frequently workers in the occupation perform the task and how important it is in their job. The frequency information is encoded in a seven-item variable group that provides information on which percentage of workers in the occupation perform the task with a given frequency, ranging from  1, ``yearly or less,'' to 7, ``hourly or more.'' For example, $34.85\%$ of \emph{Online Merchants} perform the task \emph{Receive and process payments from customers, using electronic transaction services} ``daily'' (level 5), and $23.91\%$ \emph{Calculate revenue, sales, and expenses, using financial accounting or spreadsheet software} ``more than weekly''  (level 4).

To convert these frequencies into estimates of the share of time that workers in a given occupation spend on each task, we explore two approaches. In the first (``distributive'') approach, we try to make reasonable assumptions about how much time a worker spends on tasks in a given frequency category. We list these assumptions in the third column of Table~\ref{tab:ft_weights} below. Next, we assume that workers distribute the time allotted to each frequency category and weight this by the percentage of responders of the tasks in the category. For example, according to Table~\ref{tab:ft_weights} tasks performed ``several times daily'' (level 6) are assigned a weight of 0.25. That is, we assume that tasks with level 6, taken all such tasks together, amount to $25\%$ of the working time in an occupation. If the occupation contains three tasks at frequency level 6 with weights $a\%$, $b\%$, and $c\%$, then these three tasks together, with weights $\frac{a}{a + b +c}$, $\frac{b}{a + b + c}$, and $\frac{c}{a + b + c}$, compose $25\%$ of total working time. We then multiply these weights with their distributive weight (in this case, 0.25) and repeat this process across all 7 frequency categories. This yields the final weights for each of an occupation's tasks. 

\begin{table}[htbp]
\centering
\renewcommand\cellalign{t}
\renewcommand\cellgape{\Gape[4pt]}
\begin{tabular}{cccc}
\toprule
frequency scale & category description & weights (distributive)  & weights  (relevance) \\
\midrule
1 & Yearly or less &0 & 0.5\\
2 & More than yearly &0.02 & 1\\
3 & More than monthly &0.05 & 4\\
4 & More than weekly &0.08 & 48\\
5 & Daily &0.1 & 240\\
6 & Several times daily &0.25 & 480\\
7 & Hourly or more &0.50 & 1920\\
\midrule
\bottomrule
\end{tabular}
\caption{Scales, description, and weights under the two approaches to turn frequency categories in the \emph{Task Ratings} O*NET file into duration shares.}
\label{tab:ft_weights}
\end{table}

In the second (``relevance'') approach, we instead interpret the frequency information for each task as weights that directly express the amount of time workers spend on this task. To do so, we choose a weight for each of the seven frequency categories. These weights are listed in the fourth column of Table~\ref{tab:ft_weights}. Each weight is multiplied by the worker share information. Next, these products are summed and normalized such that they add up to one within each occupation.

Finally, we need to determine how much of the time spent on each task is dedicated to programming. To estimate this, we rely on an open source large language model, Llama 3.3. We provide the model with three different prompts --- listed a the end of this section --- to arrive at three different estimates of the programming intensity of each task. Table~\ref{tab:prompts_ps_example} provides examples of tasks and the extent to which they require programming using all three prompts.

\begin{figure}[h!]
    \centering
    \includegraphics[width = \linewidth]{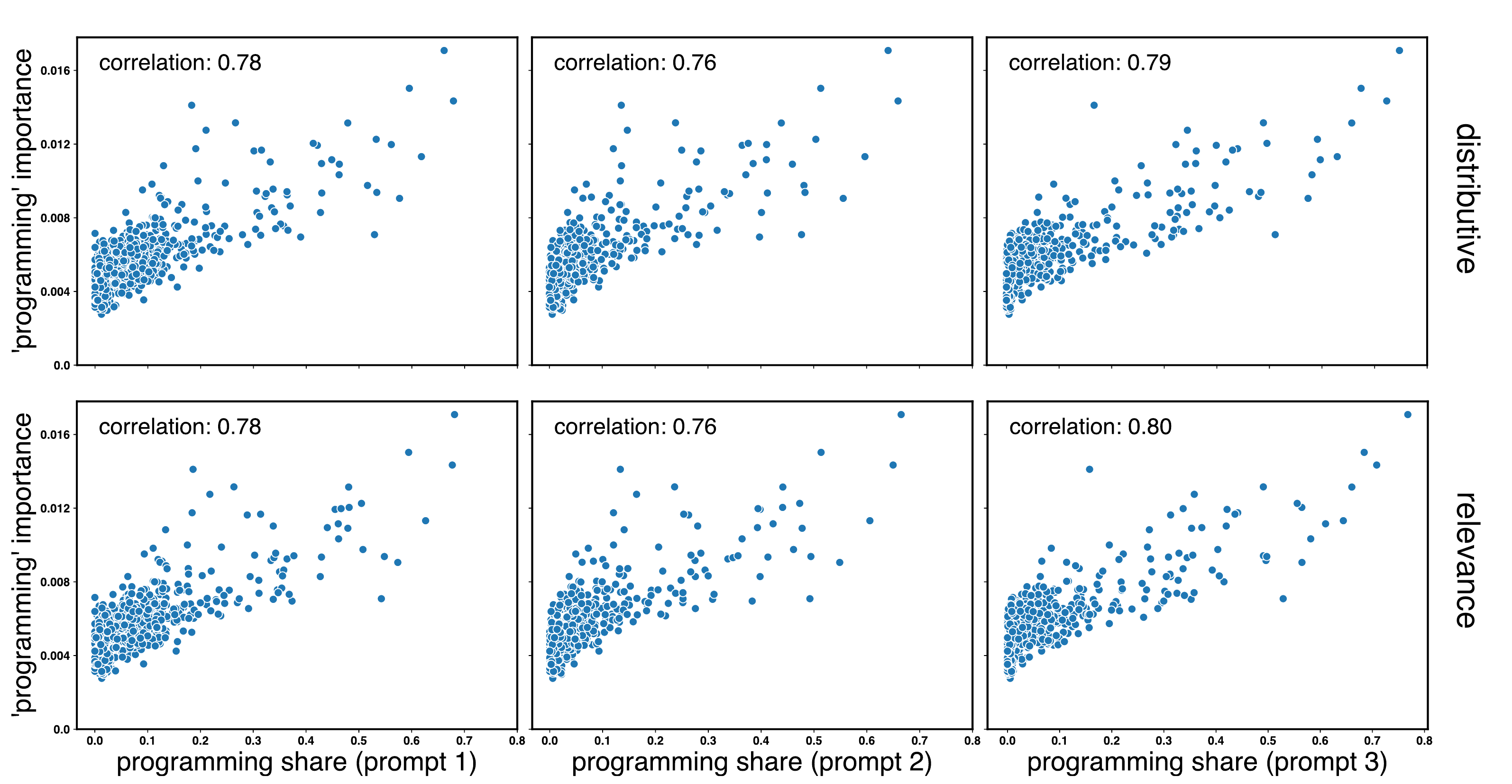}
    \caption{Programming tasks versus programming skills. Scatter plots of programming shares by occupation based on the three different prompts and the distributive (top row) or the relevance (bottom row)  approach  of turning frequencies into time shares against the importance of programming skills as reported in  O*NET.}
    \label{fig:ps_vs_im}
\end{figure}

Figure~\ref{fig:ps_vs_im}  provides scatter plots that compare our different estimates of the amount of time spent on programming tasks in each occupation to the importance of programming \emph{skills} (not tasks) for the occupation as reported in O*NET. The graph shows that all approaches yield estimates that correlate strongly with the importance of programming skill requirements, with correlations ranging between .76 and .80. In general, the correlation is highest for the third, most detailed, prompt.

To arrive at the total wage sum related to programming tasks in the US in 2023, we use two different datasets. The first uses information taken from the BLS on employment and wages:
\begin{align}
\text{Programming Wage Sum}^{BLS} =  & r\times \sum_o\text{annual wage}_o^{BLS}\times \text{employment}_o^{BLS}\times \nonumber\\ \bigl(
& \sum_{t\in \Theta_o}\text{working time}_{t,o}\times\text{programming share}_{t,o} \bigr),
\end{align}
where $\text{annual wage}_o^{BLS}$ is the average annual wage in occupation $o$ reported by the BLS, $\text{employment}_o^{BLS}$ the number of employees in occupation $o$ according to the BLS, $\text{share time}_{t,o}$ the estimated share of time that workers in occupation $o$ spend on task $t$ based on O*NET information, $\text{programming share}_{t,j}$ the LLM's estimated share of task $t$  spent on programming, and $r=1.449$ the average wage-to-compensation ratio in the US according to the BLS~\cite{bureauemployer, svanberg2024beyond} to account for wage-related costs borne by the employer.

The second dataset is the 2023 American Community Survey (ACS). The ACS contains a 1\% weighted random sample of individuals in the US. We aggregate these data to the occupation level, using the sampling weights as follows:
\begin{align}
\text{Programming Wage Sum}^{ACS} = & r\times \sum_i w_i \times \text{annual wage}_i^{ACS} \nonumber\\
\bigl(& \sum_{t\in \Theta_{o(i)}}\text{working time}_{t,o(i)}\times\text{programming share}_{t,o(i)}\bigr).
\end{align}
where $w_i$ is the frequency weight of individual $i$ in the ACS, $\text{annual wage}_i^{ACS}$ the annual salary listed for individual $i$ and $o(i)$ individual $i$'s occupation.

Unlike the BLS data, the ACS samples individuals of all ages and employment statuses (including self-employed and part-time workers). We follow prior literature~\cite{autor2013growth} to adjust top-coded annual wages and filter individuals to the active working age population. Occupations in the ACS are sometimes slightly more aggregated than in O*NET: about 110 SOC codes in ACS correspond to multiple occupation titles in O*NET. In these cases, we average the O*NET scores across the disaggregated SOC occupations that are associated with the (more aggregate) ACS occupation.

\begin{table}[htbp]
\centering
\renewcommand\cellalign{t}
\renewcommand\cellgape{\Gape[4pt]}
\begin{tabular}{ccccc}
\toprule
\ & \multicolumn{2}{c}{distributive}& \multicolumn{2}{c}{relevance} \\
\cline{2-3}\cline{4-5}
\ & BLS      & ACS    & BLS      & ACS\\
\midrule
prompt 1 & $928.33\pm 0.10$      & $1063.41\pm 0.08$    & $937.46\pm 0.09$      & $1080.72\pm 0.07$       \\
prompt 2 & $637.28\pm 0.06$      & $764.63\pm 0.06$    & $641.04\pm 0.06$      & $772.12\pm 0.06$       \\
prompt 3 & $677.91\pm 0.09$      & $760.64\pm 0.08$    & $674.32\pm 0.08$      & $763.45\pm 0.07$       \\
\midrule
\bottomrule
\end{tabular}
\caption{Estimated wage sum for programming tasks in the US in 2023 in Billions of USD. Ranges ($\pm x$) reflect 95\% simulated confidence intervals based on the uncertainty bands provided by O*NET for task frequencies. }\label{tab:programming_salary}
\end{table}

Table~\ref{tab:programming_salary} presents estimated wage sums for programming tasks in the US based on the two different samples, the three different prompts and the two different approaches to turn frequency into time-share information. Wage sums range from 637B-1,063B USD. Note that ACS based estimates always exceed BLS based results. This is because BLS only counts full-time employees and omits self-employed individuals, while the ACS samples individuals regardless of their employment status. Wage sums based on the first prompt are generally higher than those based on the other two prompts, whose estimates are very close to one another. This difference is mainly driven by tasks that are only marginally related to programming, receiving a score of 1 or 2 out of 5. Overall, the wage sums reported in Table~\ref{tab:programming_salary} imply that $2\sim 4\%$ of US GDP is spent on remunerating pure programming work.

\paragraph{Prompts to estimate the programming share of tasks.} We supply three different prompts to a Llama 3.3 model to estimate which percentage of their time workers dedicate to programming tasks. The first prompt returns a score on a scale that ranges from 1 to 5, where 1 signals that $0\%$ of working time is dedicated to programming and 5 means that almost all working time is used for programming. The detailed prompt is:

\begin{promptbox}
Analyze the relationship between a specific job task and the skill of computer programming.

**Job Role:** ``  ''

**Task Description:** ``  ''

**Instructions:**
            
1.  Consider the definition of "computer programming" as the act of writing, modifying, testing, debugging, or maintaining code or scripts (e.g., using languages like Python, Java, C++, SQL, shell scripts, PowerShell, etc.).

2.  Evaluate how much of the *specific task described above* involves performing computer programming activities. Do not evaluate the entire job role, only the task provided.

3.  Provide *only* a single numerical score from 1 to 5 based on the scale below. Do not add any explanation or text other than the score itself.

**Scoring Scale:**

* **1:** This task is not related to computer programming at all.

* **3:** Performing this task involves spending around half of the time on computer programming activities.

* **5:** This task is very related to computer programming, and performing it involves spending almost $90\%$ of the time on computer programming activities.

**Score:** 
\end{promptbox}

The second prompt is a more detailed version of the first one. For each task in each job, it returns a score on a scale that ranges from 0 to 5 where 0 stands for $0\%$ of working time dedicated to programming and 5 for almost all time dedicated to programming. The percentage range of each index is given in the prompt. The detailed prompt is:

\begin{promptbox}
Analyze the relationship between a specific job task and the skill of computer programming.

**Job Role:** ``  ''

**Task Description:** ``  ''

**Instructions:**

1.  Consider the definition of "computer programming" as the act of writing, modifying, testing, debugging, or maintaining code or scripts (e.g., using languages like Python, Java, C++, SQL, shell scripts, PowerShell, etc.).

2.  Evaluate how much of the *specific task described above* involves performing computer programming activities. Do not evaluate the entire job role, only the task provided.

3.  Provide *only* a single numerical score from 0 to 5 based on the scale below. Do not add any explanation or text other than the score itself.

**Scoring Scale (0-5):**

* **0:** **None.** The task involves absolutely no computer programming activities.
*(Estimated programming proportion: $0\%$)*
                
* **1:** **Minimal / Trace.** Programming is present but extremely limited or incidental, a negligible part of the task.
*(Estimated programming proportion: $1\%$ - $10\%$)*
                
* **2:** **Minor / Some.** Programming is a recognizable but small part of the task, clearly secondary to other activities.
*(Estimated programming proportion: $11\%$ - $25\%$)*

* **3:** **Moderate.** Programming constitutes a significant portion, but typically less than or roughly equal to other activities within the task.
*(Estimated programming proportion: $26\%$ - $50\%$)*

* **4:** **Substantial / Major.** Programming is a primary activity, taking up a clear majority of the effort for the task.
*(Estimated programming proportion: $51\%$ - $75\%$)*
                
* **5:** **Main.** Programming is the main activity, taking up all or almost all of the effort for the task.
*(Estimated programming proportion: $76\%$ - $10\%$)*

**Score:** 

\end{promptbox}

The third prompt is even  more detailed than the first two. For each task in each job, it returns a score on a scale that ranges from 0 to 100 that indicates the programming related working time percentage. Several examples are provided to anchor the scale. The detailed prompt is:
\begin{promptbox}
**Prompt for Llama 3:**

**Context:**

You are an AI assistant tasked with analyzing job roles and specific tasks within those roles to estimate the proportion of time dedicated to programming activities. I will provide you with a job title and a description of a single task performed within that job.

**Your Goal:**

Based on the provided job title and task description, estimate the approximate percentage of time that would be spent **actively writing, testing, debugging, or deploying code** for this specific task. Your output should be **only the numerical percentage value**. Focus on inferring the nature of the work from the overall description, rather than relying solely on specific keywords.

**Input:**

* **Job Title:** ``

* **Task Description:** ``

**Instructions for Your Analysis (Internal Thought Process - Do Not Include in Output):**

1.  **Holistic Task Understanding in Job Context:**

* Read the `Task Description` thoroughly. Instead of just looking for keywords like "develop" or "code," try to understand the overall objective and the types of activities implicitly required to achieve it, considering the typical responsibilities of the given `Job Title`.

* For instance, a task described as "resolve customer-reported performance bottlenecks in the data processing pipeline" implies deep investigation, potentially code profiling, optimization, and testing, even if the word "coding" isn't explicitly used.

2.  **Infer Programming-Related Activities:**

* Based on your holistic understanding, determine what proportion of the task likely involves direct engagement with programming activities (e.g., designing algorithms that will be coded, writing new code, modifying existing code, scripting, debugging complex systems, implementing tests, or managing code deployment).

* Consider the full software development lifecycle if implied by the task.

3.  **Consider Implied Non-Programming Activities:**

* Also, identify parts of the task that, based on its nature, would likely involve significant non-programming activities. This could include extensive research before any coding can begin, detailed planning and architecting, writing documentation, attending meetings for coordination, user interviews, data gathering and analysis (if not directly scripting it), or system monitoring and analysis that doesn't immediately lead to code changes.

4.  **Estimate the Percentage based on Inferred Effort:**

* Determine a single numerical percentage representing your best estimate of the time spent on direct programming activities for this *specific task*, based on the inferred balance of efforts.

**Output Requirement:**

* Return **ONLY the numerical percentage value**. For example, if you estimate $30\%$, output only `30`. Do not include the '$\%$' symbol or any other explanatory text.

**Examples of Task Analysis (Illustrative - For your understanding of the analysis process only, not the output format for the actual task. Note how the reasoning infers activities):**

* **Example 1 (Low Programming):**

* **Job Title (Example):** Software Engineer

* **Task Description (Example):** "Collaborate with the product team to define specifications for a new user authentication module."

* **Internal Estimation Logic (Example):** The description emphasizes collaboration ("collaborate") and definition of requirements ("define specifications"). This strongly suggests activities like discussion, documentation, and planning, which are primarily pre-coding. *This specific task* is about laying the groundwork. Estimated programming time for *this specific task*: $10\%$.

* **Example 2 (Mid-Range Programming):**

* **Job Title (Example):** Data Scientist

* **Task Description (Example):** "Investigate anomalies in sales data and present findings to the marketing department."

* **Internal Estimation Logic (Example):** "Investigate anomalies" might involve some scripting for data extraction and initial analysis. However, "present findings" implies data interpretation, visualization, report preparation, and communication. Estimated programming time for *this specific task*: $35\%$.

* **Example 3 (Mid-Range Programming):**

* **Job Title (Example):** DevOps Engineer

* **Task Description (Example):** "Oversee the migration of our primary application servers to a new cloud provider, ensuring minimal downtime and performance continuity."

* **Internal Estimation Logic (Example):** "Oversee the migration" involves planning and coordination. While automation scripts (programming) will be part of ensuring "minimal downtime and performance continuity," a significant portion involves project management and validation. Estimated programming time for *this specific task*: $40\%$.

* **Example 4 (Low Programming):**

* **Job Title (Example):** UI/UX Designer

* **Task Description (Example):** "Create interactive prototypes for the upcoming mobile application redesign based on user feedback and usability testing results."

* **Internal Estimation Logic (Example):** Prototyping here focuses on design tools and user experience demonstration, not general-purpose programming, even if some tools have coding-like features. Estimated programming time for *this specific task*: $15\%$.

* **Example 5 (High Programming):**

* **Job Title (Example):** Full Stack Developer

* **Task Description (Example):** "Refactor the existing monolithic backend service into a set of microservices to improve scalability and maintainability."

* **Internal Estimation Logic (Example):** "Refactor... into a set of microservices" is a substantial software engineering effort involving deep code analysis, writing significant new code, and extensive testing. Estimated programming time for *this specific task*: $80\%$.

* **Example 6 (Very High Programming):**
* **Job Title (Example):** Computer Programmers

* **Task Description (Example):** "Perform or direct revision, repair, or expansion of existing programs to increase operating efficiency or adapt to new requirements."

* **Internal Estimation Logic (Example):** This task is the core of what a computer programmer does. "Revision, repair, or expansion of existing programs" directly translates to reading, understanding, modifying, testing, and debugging code. Estimated programming time for *this specific task*: $95\%$.

* **Example 7 (Very High Programming):**

* **Job Title (Example):** Computer Programmers

* **Task Description (Example):** "Write, update, and maintain computer programs or software packages to handle specific jobs such as tracking inventory, storing or retrieving data, or controlling other equipment."

* **Internal Estimation Logic (Example):** "Write, update, and maintain computer programs" is unequivocally direct programming work. This involves the full cycle of coding for specific functionalities. Estimated programming time for *this specific task*: $95\%$.

* **Example 8 (Very High Programming):**

* **Job Title (Example):** Web Developers

* **Task Description (Example):** "Write supporting code for Web applications or Web sites."

* **Internal Estimation Logic (Example):** "Write supporting code" is a direct statement of programming activity within the context of web development (e.g., backend logic, frontend scripting, API integration). Estimated programming time for *this specific task*: $90\%$.

* **Example 9 (Very High Programming):**

* **Job Title (Example):** Bioinformatics Technicians

* **Task Description (Example):** "Write computer programs or scripts to be used in querying databases."

* **Internal Estimation Logic (Example):** "Write computer programs or scripts" for database querying is a clear programming task, essential for data retrieval and analysis in bioinformatics. Estimated programming time for *this specific task*: $90\%$.

* **Example 10 (Very High Programming):**

* **Job Title (Example):** Atmospheric and Space Scientists

* **Task Description (Example):** "Develop computer programs to collect meteorological data or to present meteorological information."

* **Internal Estimation Logic (Example):** "Develop computer programs" for data collection or presentation directly points to software development, likely involving data processing, numerical modeling, or visualization coding. Estimated programming time for *this specific task*: $85\%$ (allowing for some potential non-coding research or data interpretation elements within the broader task).

* **Example 11 (Very Low Programming):**

* **Job Title (Example):** Chief Executives

* **Task Description (Example):** "Appoint department heads or managers and assign or delegate responsibilities to them."

* **Internal Estimation Logic (Example):** This task is purely managerial and strategic, involving decision-making, leadership, and organizational structuring. There is no implied programming. Estimated programming time for *this specific task*: $0\%$.

* **Example 12 (Very Low Programming):**

* **Job Title (Example):** Education and Childcare Administrators, Preschool and Daycare

* **Task Description (Example):** "Teach classes or courses or provide direct care to children."

* **Internal Estimation Logic (Example):** This task involves direct pedagogical activities, caregiving, and interpersonal interaction, with no programming component. Estimated programming time for *this specific task*: $0\%$.

* **Example 13 (Very Low Programming):**

* **Job Title (Example):** Food Service Managers

* **Task Description (Example):** "Test cooked food by tasting and smelling it to ensure palatability and flavor conformity."

* **Internal Estimation Logic (Example):** This task involves sensory evaluation and quality control related to food, entirely non-programming. Estimated programming time for *this specific task*: $0\%$.

* **Example 14 (Very Low Programming):**

* **Job Title (Example):** Gambling Managers

* **Task Description (Example):** "Notify board attendants of table vacancies so that waiting patrons can play."

* **Internal Estimation Logic (Example):** This task is operational and communicative, focusing on managing customer flow and staff coordination within a gambling establishment. No programming is involved. Estimated programming time for *this specific task*: $0\%$.

* **Example 15 (Very Low Programming):**

* **Job Title (Example):** Postmasters and Mail Superintendents

* **Task Description (Example):** "Select and train postmasters and managers of associate postal units."

* **Internal Estimation Logic (Example):** This task is focused on human resources, management, and training, with no direct programming activities. Estimated programming time for *this specific task*: $0\%$.

**Now, please analyze the following and provide ONLY the numerical percentage as output, focusing on an inference from the overall description:**

* **Job Title:** ``

* **Task Description:** ``

**Begin Analysis and Provide Only the Numerical Percentage. Please only give me the numerical percentage and do not output any other text.**
\end{promptbox}

Table~\ref{tab:prompts_ps_example} provides examples of tasks and their scores from the three prompts.

\renewcommand\tabularxcolumn[1]{m{#1}} % Optional: vertical centering in X columns

\begin{table}[htbp]
\scriptsize
\renewcommand{\arraystretch}{0.9}
\caption{Fifteen example tasks and jobs from O*NET with their programming share score and corresponding percentage provided by Llama 3.3 using three different prompts.}
\label{tab:prompts_ps_example}
\begin{tabularx}{\textwidth}{X X ccc}
\toprule
\textbf{Job} & \textbf{Task} & \textbf{Prompt 1} & \textbf{Prompt 2} & \textbf{Prompt 3} \\
\midrule
Robotics Engineers & Write algorithms or programming code for ad hoc robotic applications. & 5 (87.5\%) & 5 (88\%) & 95 (95\%) \\
Biostatisticians & Write program code to analyze data with statistical analysis software. & 5 (87.5\%) & 5 (88\%) & 90 (90\%) \\
Computer Systems Engineers/Architects & Develop efficient and effective system controllers. & 5 (87.5\%) & 5 (88\%) & 80 (80\%) \\
Computer and Information Research Scientists & Analyze problems to develop solutions involving computer hardware and software. & 4 (62.5\%) & 4 (63\%) & 80 (80\%) \\
Financial Quantitative Analysts & Devise or apply independent models or tools to help verify results of analytical systems. & 4 (62.5\%) & 4 (63\%) & 70 (70\%) \\
Computer Network Architects & Develop or recommend network security measures, such as firewalls, network security audits, or automated security probes. & 4 (62.5\%) & 3 (38\%) & 60 (60\%) \\
Computer Network Architects & Design, build, or operate equipment configuration prototypes, including network hardware, software, servers, or server operation systems. & 3 (37.5\%) & 3 (38\%) & 60 (60\%) \\
Telecommunications Engineering Specialists & Implement system renovation projects in collaboration with technical staff, engineering consultants, installers, and vendors. & 2 (12.5\%) & 2 (18\%) & 40 (40\%) \\
Logistics Engineers & Develop or document reverse logistics management processes to ensure maximal efficiency of product recycling, reuse, or final disposal. & 2 (12.5\%) & 2 (18\%) & 20 (20\%) \\
Bookkeeping, Accounting, and Auditing Clerks & Reconcile records of bank transactions. & 2 (12.5\%) & 1 (5.5\%) & 10 (10\%) \\
Pump Operators, Except Wellhead Pumpers & Record operating data such as products and quantities pumped, stocks used, gauging results, and operating times. & 2 (12.5\%) & 1 (5.5\%) & 5 (5\%) \\
Operating Engineers and Other Construction Equipment Operators & Locate underground services, such as pipes or wires, prior to beginning work. & 2 (12.5\%) & 1 (5.5\%) & 5 (5\%) \\
Administrative Services Managers & Set goals and deadlines for the department. & 1 (0\%) & 0 (0\%) & 0 (0\%) \\
Cement Masons and Concrete Finishers & Spread roofing paper on surface of foundation, and spread concrete onto roofing paper with trowel to form terrazzo base. & 1 (0\%) & 0 (0\%) & 0 (0\%) \\
First-Line Supervisors of Farming, Fishing, and Forestry Workers & Treat animal illnesses or injuries, following experience or instructions of veterinarians. & 1 (0\%) & 0 (0\%) & 0 (0\%) \\
\bottomrule
\end{tabularx}
\end{table}

\section{General equilibrium effects}\label{sec:SI_geneq}

If genAI increases the productivity of programmers, this will not only affect the quantity of code produced, but also its price. Although a full fledged calibrated general equilibrium model is beyond the scope of the current paper, in this section we try to put some plausible bounds on such general equilibrium effects of genAI. To do so, we consider two scenarios. Both scenarios feature a ---for simplicity, linear---standard downward-sloping demand curve. They differ in the assumptions about the supply curve, representing polar opposite cases: perfectly elastic and perfectly inelastic supply. 

\subsection*{Scenario 1: perfectly elastic supply of code}

In this scenario, the supply of code is perfectly elastic. That is, programmers supply any quantity of code at price $p_1$ . 
For instance, if there were a large pool of identical programmers who can freely enter or exit the market for code and who all have the same outside option offering a reservation wage, $w^*$, per day, programmers  enter (or exit) the market, until the equilibrium price for code, $p^*$ is such that their earnings $qp^* = w^*$, where $q$ is the volume of code a programmer can write in a day. 

\begin{figure}[H]
    \centering
            \includegraphics[width=.8\linewidth]{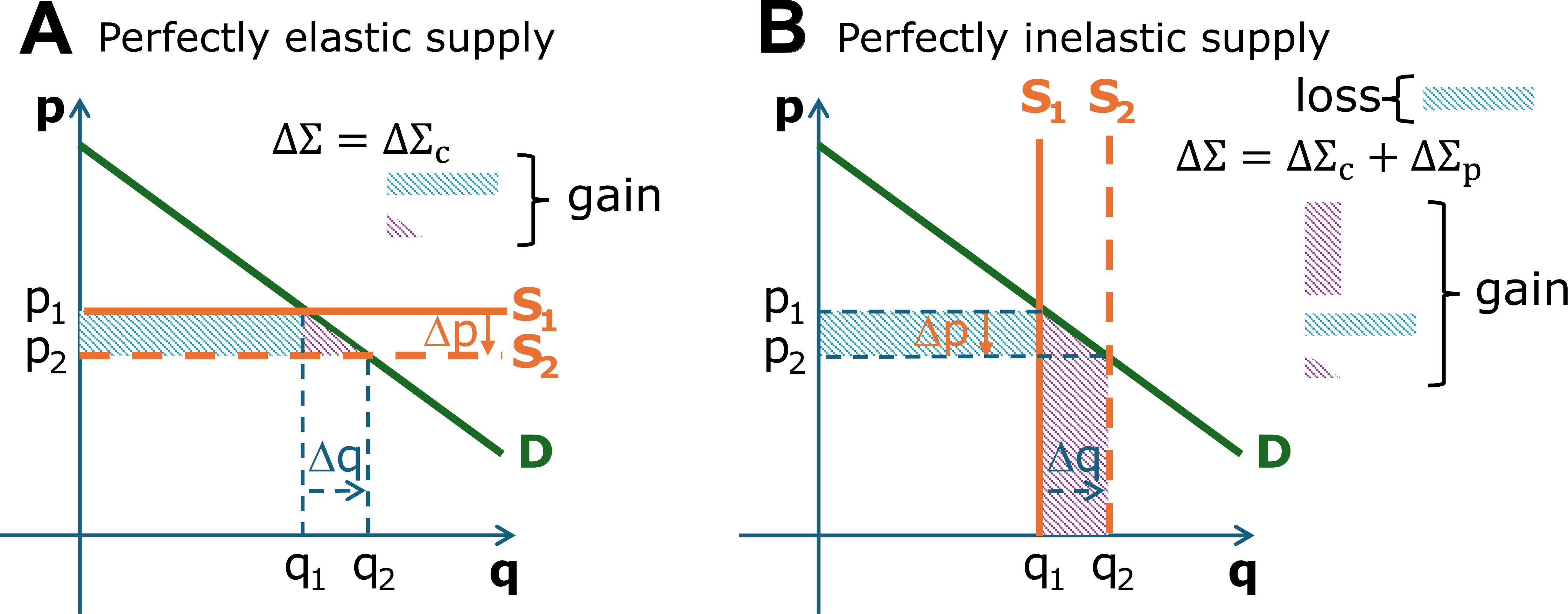}
            \par
    \caption{Changes in social surplus due to genAI. Green lines display  demand curves, orange lines supply curves in the price ($p$) - quantity ($q$) plane. With the introduction of genAI, there is a shift in supply from $S_1$ to $S_2$. The colored areas represent changes in the social surplus ($\Delta \Sigma$), consisting of changes in the producer surplus ($\Delta \Sigma_p$) and in the consumer surplus  ($\Delta \Sigma_c$). Panel \textbf{A} shows how genAI changes prices and quantities of code in a scenario where the supply of code is perfectly elastic. Panel \textbf{B} shows the same when the supply of code is perfectly inelastic.}
    \label{fig:SI_gen_eq}

\end{figure}

Under such a scenario, illustrated in Fig.~\ref{fig:SI_gen_eq}A, changes in demand will be absorbed by programmers entering or leaving the coding market without changing the price of code. 
Productivity effects of genAI now increase how much code a programmer can produce per day, shifting the supply curve down. The excess wages will attract additional programmers into the market, increasing the supply of code while moving down along the demand curve until the new equilibrium price $p_2$ is reached and incomes again match reservation wages. 

The result is an increase in the consumer surplus, $\Sigma_c$, reflecting  two effects: First, the price of code drops, leading to savings for all original consumers of code (blue rectangle). Second, at the new price, $p_2$, consumers will demand additional code, some of which will be sold below the consumer's willingness-to-pay (purple triangle). At a perfectly elastic supply, the producer surplus, $\Sigma_p$, is zero, both before and after the arrival of genAI. Changes in the total social surplus, $\Delta \Sigma = \Delta\Sigma_c + \Delta\Sigma_p$, therefore fully consist of changes in the consumer surplus.

To calculate the value of $\Delta \Sigma$, we introduce some notation:
\begin{itemize}
    \item $p_1$: price of a unit of code before introduction of genAI 
    \item $p_2$: price of a unit of code after introduction of genAI 
    \item $q_1$ : quantity of code before introduction of genAI 
    \item $q_2$: quantity of code after introduction of genAI 
    \item $L_1$: number of coders before introduction of genAI 
    \item $L_2$: number of coders after introduction of genAI 
    \item $\delta =\frac{\Delta q}{q_1} =\frac{q_2 -q_1}{q_1}$: effect of genAI, the percentage increase in quantity of code due to the adoption of genAI  
    \item $w^*$: reservation wage satisfying $w^*=\frac{p_1q_1}{L_1} =\frac{p_2q_2}{L_2}$ 
    \item $V_1 =p_1q_1$: value of code before introduction of genAI 
    \item $\eta=\frac{\Delta q/q_1}{\Delta p/p_1}$: demand elasticity of code, implying $\Delta p=\frac{\Delta q/q_1}{\eta/p_1}$ 
\end{itemize}

The change in social surplus, $\Delta \Sigma$,   equals the area of two colored areas:
\begin{align*}%\label{eq:SI_socsurplus_1}
    \Delta \Sigma^\texttt{elas} &=(p_1-p_2)\frac{q_1+q_2}{2} \\
    &= -\Delta p \frac{2q_1+\Delta q}{2} \\
    &=- \frac{\Delta q/q_1}{\eta /p_1} q_1 - \frac{\Delta q/q_1}{\eta /p_1}\frac{\Delta q}{2} \\
    &=-\frac{\Delta q}{q_1}\frac{p_1q_1}{\eta}\left( 1+\frac{1}{2}\frac{\Delta q}{q_1}\right) \\
    &=-\frac{\delta}{\eta}V_1\left(1+\frac{\delta}{2}\right)    
\end{align*}

Note that, except for the demand elasticity, $\eta$, all parameters in this expression are known.

\subsection{Scenario 2: perfectly inelastic supply of code}\label{sec:SI_inelas}

In the second scenario, supply of code is perfectly inelastic, yielding a vertical supply curve. This is a likely short-run scenario, where the labor market has insufficient time to adjust to the productivity effects of genAI. Under this scenario,  the set of programmers in the market is fixed, because it is hard to move in or out of the coding labor market. As a result, they will supply their labor at any price the market supports. 

With a fixed set of coders, productivity effects of genAI translate one-to-one into an increase in the equilibrium amount of code supplied to the market, shifting the supply curve of code  to the right. This results in changes in both consumer and producer surplus. Producers of code receive compensation for the increased volume of code. However, they also bear a cost because they receive a lower price for their code. In Fig.~\ref{fig:SI_gen_eq}B, this is represented by the two colored rectangles. The producer surplus is reduced by the area of the blue shape but increased --- because of the higher demand for code at price $p_2$  and producers’ willingness to supply this code at any price –-- by the area of the orange shape. Consequently, the net effect on the producer surplus is ambiguous. 

By contrast, consumers unambiguously gain, because the equilibrium price of code falls. As  in Scenario 1, this induces them to consume more code. As before, the increase in consumer surplus therefore equals the sum of the areas of the blue rectangle and the purple triangle. 

Taken changes in consumer and producer surplus together, the net change in social surplus is unambiguously positive and equal to the areas of the orange rectangle and the purple triangle:
\begin{align*}%\label{eq:SI_socsurplus_2}
    \Delta \Sigma^\texttt{inelas} &=\Delta q\frac{p_1+p_2}{2} \\
    &= \Delta q p_1 + \frac{1}{2} \Delta q \Delta p \\
    &=\Delta q p_1 + \frac{1}{2} \Delta q \frac{\Delta q/q_1}{\eta / p_1} \\
    &=\frac{\Delta q}{q_1}p_1 q_1\left(1+\frac{1}{2}\frac{\Delta q}{q_1} \frac{1}{\eta} \right) \\
    &=\delta V_1\left(1+\frac{1}{2}\frac{\delta}{\eta}\right)    
\end{align*}

\subsection{Comparing changes in social surplus}

To calculate the changes in social surplus in the two scenarios sketched above, we need an estimate of $\eta$, the elasticity of the demand of code.  Software code is often tailored to a specific use case and therefore will typically have few close substitutes. Therefore, we expect demand to be relatively insensitive to price changes, suggesting inelastic demand: $\eta<-1$. 

We do not know of any estimates of the elasticity of demand for software code. To get a sense of its likely magnitude, we can look at other types of intellectual property (IP) that are similarly high-value, intangible and with few close substitutes: patents and trademarks. For IP protected by patents  \cite{rassenfosse2012price} estimate an elasticity of $-0.3$, whereas \cite{rassenfosse2021price} estimates an elasticity of between $-0.25$ and $-0.40$ for IP protected by trademarks. Combining the $\eta = -0.3$ estimate for patent-based IP with the average of our range for the coding-related wage sum in the US of $V_1 = 787$ billion USD and our baseline estimates for the effect on productivity effect of genAI in the US at the end of 2024 of $\delta = e^{\beta_{base} * 29\%}-1 = 3.6\% $ adoption rate, we arrive at the following estimates for the increases in the social surplus in billions of USD:
\begin{itemize}
    \item $\Sigma^\texttt{elas} = -\frac{\delta}{\eta}V_1\left(1+\frac{\delta}{2}\right)= \frac{0.035}{0.3}*787\left(1+\frac{0.035}{2}\right)=93.1$
    \item $\Sigma^\texttt{inelas} = \delta V_1\left(1+\frac{1}{2}\frac{\delta}{\eta}\right) = 0.035*787\left(1-\frac{1}{2}\frac{0.035}{0.3}\right)=25.9  $ 
\end{itemize}
Given this range of possible long-run outcomes, our initial, short-run, estimate of 27.6 billion USD (using again the average estimate of the wage sum) is relatively conservative.

\clearpage 
\bibliographystyle{unsrt}

\bibliography{main}

\end{document}